\documentclass[preprint2]{aastex6}

\usepackage{natbib}
\shorttitle{Molecular-Cloud-Scale Chemical Composition II: W3(OH)}
\shortauthors{Nishimura et al.}

\begin{document}

\title{Molecular-Cloud-Scale Chemical Composition II: \\
       Mapping Spectral Line Survey toward W3(OH) in the 3 mm Band}

\author{Yuri Nishimura\altaffilmark{1,2,3},
        Yoshimasa Watanabe\altaffilmark{4,3},
        Nanase Harada\altaffilmark{5},
        Takashi Shimonishi\altaffilmark{6,7},\\
        Nami Sakai\altaffilmark{7},
        Yuri Aikawa\altaffilmark{9},
        Akiko Kawamura\altaffilmark{2},
        and 
        Satoshi Yamamoto\altaffilmark{3,10}}

\altaffiltext{1}{Institute of Astronomy, The University of Tokyo, 
                 2-21-1, Osawa, Mitaka, Tokyo 181-0015, Japan}
\altaffiltext{2}{Chile Observatory, National Astronomical Observatory of Japan, 
                 2-21-1, Osawa, Mitaka, Tokyo 181-8588, Japan}
\altaffiltext{3}{Department of Physics, The University of Tokyo, 
                 7-3-1, Hongo, Bunkyo-ku, Tokyo 113-0033, Japan}
\altaffiltext{4}{Faculty of Pure and Applied Sciences, University of Tsukuba, 
                 1-1-1, Tennodai, Tsukuba, Ibaraki 305-8577, Japan}
\altaffiltext{5}{Academia Sinica Institute of Astronomy and Astrophysics, 
                 No.1, Sec. 4, Roosevelt Rd, 10617 Taipei, R.O.C., Taiwan}                 
\altaffiltext{6}{Frontier Research Institute for 
                 Interdisciplinary Sciences, Tohoku University, 
                 Aramakiazaaoba 6-3, Aoba-ku, Sendai, Miyagi 980-8578, Japan}
\altaffiltext{7}{Astronomical Institute, Tohoku University, 
                 Aramakiazaaoba 6-3, Aoba-ku, Sendai, Miyagi 980-8578, Japan}
\altaffiltext{8}{RIKEN, 2-1 Hirosawa, Wako, Saitama 351-0198, Japan}
\altaffiltext{9}{Department of Astronomy, The University of Tokyo, 
                 7-3-1, Hongo, Bunkyo-ku, Tokyo 113-0033, Japan}
\altaffiltext{10}{Research Center for the Early Universe, The University of Tokyo, 
                  7-3-1, Hongo, Bunkyo-ku, Tokyo, 113-0033, Japan}

\begin{abstract}
In order to study a molecular-cloud-scale chemical composition, 
we have conducted a mapping spectral line survey 
toward the Galactic molecular cloud W3(OH), 
which is one of the most active star forming regions in the Perseus arm, 
with the NRO 45 m telescope. 
We have observed the area of $16'\times16'$, 
which corresponds to 9.0 pc $\times$ 9.0 pc. 
The observed frequency ranges are $87-91$, $96-103$, and $108-112$ GHz. 
We have prepared the spectrum averaged over the observed area, 
in which 8 molecular species CCH, HCN, HCO$^+$, HNC, CS, SO, C$^{18}$O, 
and $^{13}$CO are identified. 
On the other hand, the spectrum of the W3(OH) hot core 
observed at a 0.17 pc resolution 
shows the lines of various molecules 
such as OCS, H$_2$CS CH$_3$CCH, and CH$_3$CN, 
in addition to the above species. 
In the spatially averaged spectrum, 
emission of the species concentrated just around the star-forming core 
such as CH$_3$OH and HC$_3$N is fainter than in the hot core spectrum, 
whereas emission of the species widely extended over the cloud 
such as CCH is relatively brighter. 
We have classified the observed area into 5 subregions 
according to the integrated intensity of $^{13}$CO, and have evaluated 
the contribution to the averaged spectrum from each subregion. 
The CCH, HCN, HCO$^+$, and CS lines can be seen 
even in the spectrum of the subregion 
with the lowest $^{13}$CO integrated intensity range ($< 10$ K km s$^{-1}$). 
Thus, the contributions of the spatially extended emission is 
confirmed to be dominant in the spatially averaged spectrum. 
\end{abstract}


\section{Introduction} \label{introduction}

\begin{figure*}
\begin{center}
\includegraphics[width=\hsize, bb=0 0 720 288]{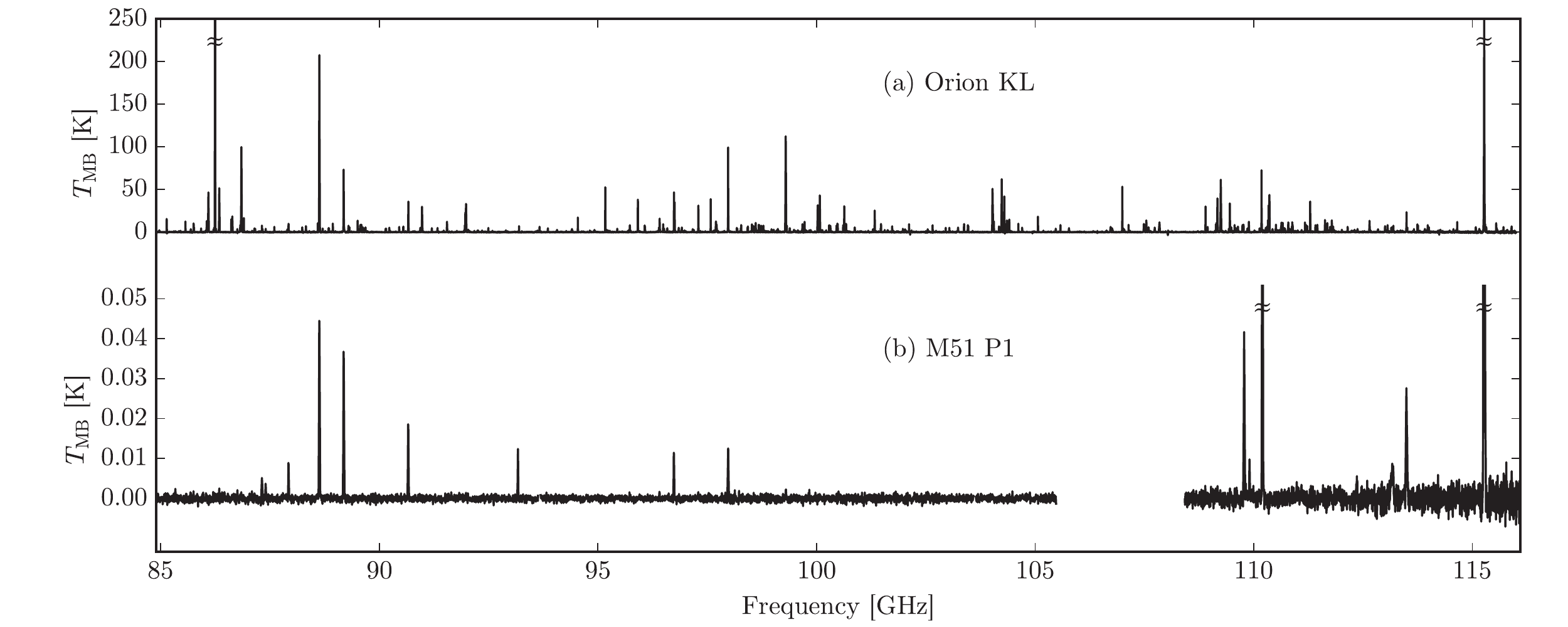}
\caption{The 3 mm band spectra of (a) galactic high mass star-forming regions 
Orion KL \citep[][0.04 pc resolution]{watanabe2015spectral}, 
and (b) external galaxies M51 P1 \citep[][1 kpc resolution]{watanabe2014spectral}. 
\label{Gal_extragal}}
\end{center}
\end{figure*}

Recently, chemical compositions of nearby extragalactic objects 
have been studied by taking advantage of increased sensitivity 
of single-dish radio telescopes and radio interferometers. 
Unbiased spectral line surveys at millimeter wavelengths 
have been conducted toward a variety of targets; 
active galactic nuclei \citep[AGNs; e.g.,][]{aladro2013lambda}, 
starburst galaxies \citep[e.g.,][]{martin2006M82}, 
and (ultra) luminous infrared galaxies 
\citep[(U)LIRGs; e.g.,][]{martin2011Arp220}, 
as well as low-metallicity dwarf galaxies 
\citep[e.g.,][]{nishimura2016IC10, nishimura2016LMC}. 
Such observations have been carried out 
not only for nuclear regions of nearby galaxies, 
but also for positions in normal spiral arms 
\citep[e.g.,][]{watanabe2014spectral}. 
The chemical composition of each galaxy revealed by 
the spectral line survey in these studies is discussed in relation to 
a galaxy type, an evolutional stage, and a physical environment. 
Chemical composition will be a powerful probe 
for star-forming activities in external galaxies, 
if we could link it to astrochemical concepts established 
in nearby star-forming regions in our Galaxy. 
Yet, comparison between extragalactic and galactic observations 
requires special attention to the large difference of the size 
that can be resolved in their observations. 
Due to the limited spatial resolution, 
these extragalactic studies are of a molecular-cloud scale (several ten pc), 
while we can readily resolve individual dense cores ($<0.1$ pc) 
in molecular clouds for Galactic sources. 
Figure \ref{Gal_extragal} shows the 3 mm-band spectra of 
the Galactic high-mass star-forming region (Orion KL) 
and the external galaxy (spiral arm of M51). 
The spectral intensity patterns are different between them. 
However, it is not obvious at a glance 
which molecular-line feature in the extragalactic spectrum 
reflects the internal star-forming activities. 
Emission from extended regions other than dense star-forming cores 
would contribute to the extragalactic spectrum 
at a molecular-cloud scale. 
For appropriate interpretation of extragalactic spectra 
in terms of star formation activities, it is very important 
to know the contribution of the extended component for each molecular line. 
Indeed, importance of such works is now being recognized 
\citep[e.g.,][]{kauffmann2017molecular}. 

To fill the spatial resolution gap 
between galactic and extragalactic observations 
and understand the meaning of the molecular-cloud-scale chemical composition, 
the mapping spectral line survey observation 
toward a Galactic cloud is useful. 
By averaging the spectra over the large mapped area 
covering a whole molecular cloud, 
we can simulate the spectrum of Milky Way 
just as we observe it from the other galaxy. 
The mapping line survey will also provide us 
with a deeper insight into the origin of 
the molecular-cloud-scale chemical composition, 
because we can also study spatial distributions 
of molecules within a cloud. 
Similar mapping observations of various molecular species 
in the Galactic molecular clouds have recently been reported for 
Orion B \citep{pety2017anatomy} and W49A \citep{nagy2015physical}, 
in the 84--116 GHz range and 330--373 GHz range, respectively. 

This is the second paper of a series of our work 
on the mapping spectral line survey of molecular clouds. 
In the first paper \citep{watanabe2017W51}, 
we conducted the mapping spectral line survey observation 
toward the active star-forming region W51. 
W51 is located near the tangent point 
of the Sagittarius arm \citep{sato2010trigonometric}. 
We revealed the molecular-cloud-scale chemical composition of W51, 
and found that it reflects a relatively diffuse part of molecular clouds 
rather than dense star-forming cores. 
In this study, we extend our work to another molecular cloud W3(OH). 
W3(OH) is a massive star-forming region in the Perseus arm, 
which is located in the outer part of our Galaxy. 
The Galactocentric distance is 9.95 kpc \citep{xu2006distance}, 
and the distance from the sun is 1.95 kpc. 
By comparing W3(OH) with W51, 
we can also examine the environmental effect on the molecular-cloud-scale 
chemical composition. 

\begin{figure}
\begin{center}
\includegraphics[width=\hsize, bb=0 0 567 535]{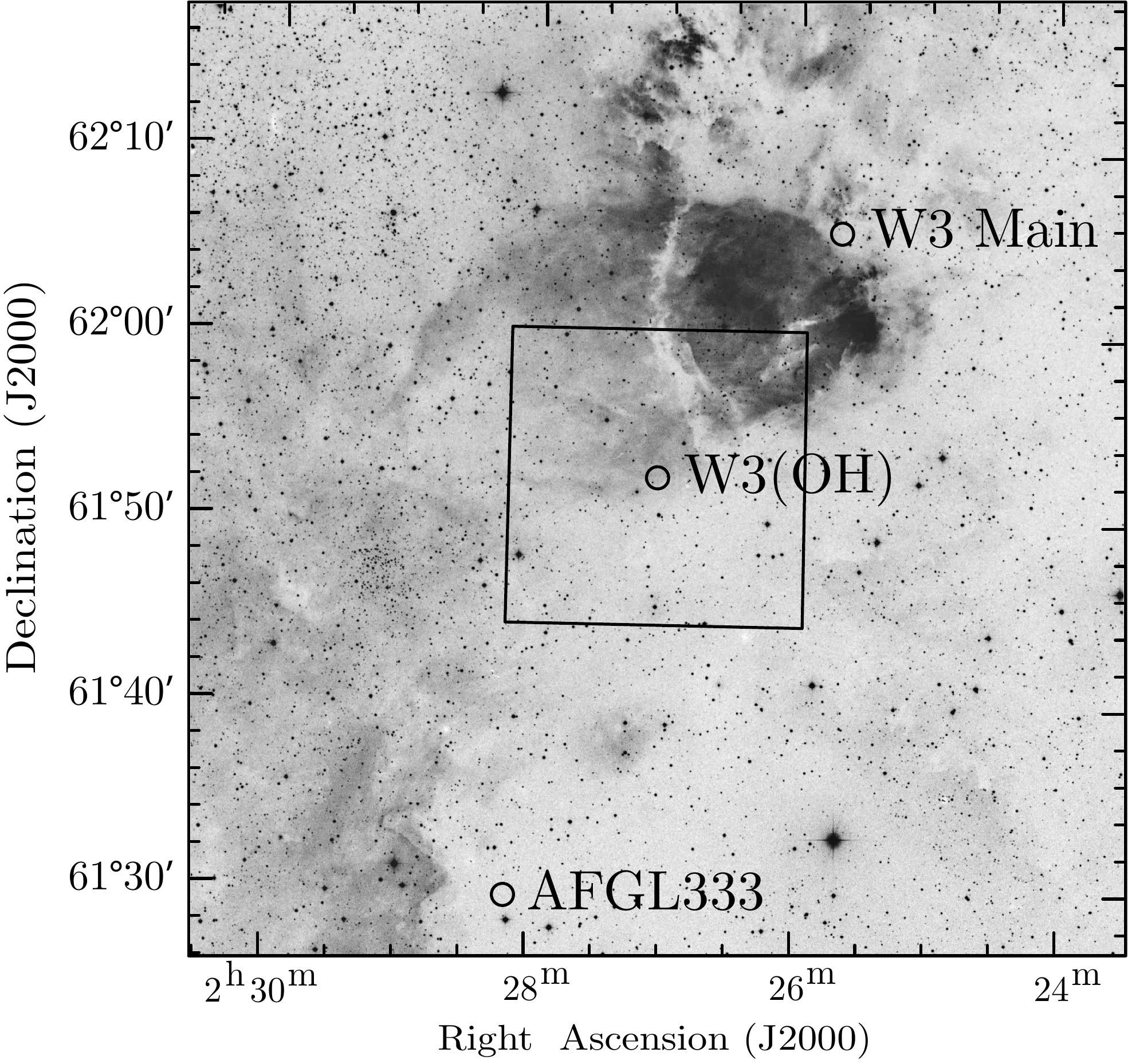}
\caption{The Digitized Sky Survey image (R band; 700 nm) of the W3 region. 
Three star-forming clouds, W3 Main, W3(OH), and AFGL333, are indicated by circles. 
We observed the area enclosed by a solid black line. 
The Digitized Sky Survey was produced at the Space Telescope Science Institute 
and the images of these surveys are based on photographic data obtained 
using the Oschin Schmidt Telescope on Palomar Mountain 
and the UK Schmidt Telescope. 
\label{optical}}
\end{center}
\end{figure}

In the W3 giant molecular cloud, the most active star-forming regions are 
W3 Main, W3(OH), and AFGL333 (Figure \ref{optical}). 
These three star-forming regions are located in the eastern edge 
of the W3 giant molecular cloud. 
This structure is considered to be formed probably from materials 
swept up by the expansion of the W4 \ion{H}{2} region \citep{lada1978W3}. 
W3(OH), the target of this mapping line survey, is an ultracompact \ion{H}{2} region 
harboring many OH maser spots \citep{wynnwilliams1972W3} 
and the second active star-forming region after W3 Main. 
A size of the hot core is estimated to be $0.02\times0.01$ pc 
\citep{wyrowski1999hot}. 

The W3 giant molecular cloud has been studied in various wavelengths 
in order to investigate its star-formation activities and evolutional stages. 
Infrared maps (70 $\mu$m, 160 $\mu$m, 250 $\mu$m, 350 $\mu$m, and 500 $\mu$m) 
were obtained with \emph{Herschel} \citep{rivera-ingraham2013W3}. 
The distributions of CO and atomic carbon were also reported 
\citep[e.g.,][]{sakai2006atomic}. 
The spectral line survey from 85 to 115 GHz was conducted 
toward several star-forming cores 
in the W3 giant molecular cloud \citep{kim2006W3}. 
Hence, the W3 giant molecular cloud is regarded as 
one of the most favorable targets for molecular-cloud-scale chemical studies. 

In this study, we present the result of our mapping observation toward W3(OH). 
Here, we focus on the molecular-cloud-scale chemical composition 
rather than the properties of each star-forming region. 
The rest of this paper is organized as follows. 
In Section \ref{observation}, we describe the observations and data reduction. 
In Section \ref{results}, 
we show the observed spectra and line profiles, 
and present the distribution of key molecular species. 
Data analysis and discussions, 
including the correlation of each molecular species 
with $^{13}$CO and HCO$^+$ 
and interpretations of the spatially averaged spectrum, 
are presented in Section \ref{discussion}. 
We also compare the spectrum of W3(OH) with those of other sources. 
Finally, we summarize main conclusions in Section \ref{summary}.

\section{Observations and data reduction} \label{observation}

The observations were carried out with 
the 45 m radio telescope at Nobeyama Radio Observatory (NRO) in March 2015. 
We observed the frequency ranges of $87-91$, $96-103$, and $108-112$ GHz. 
The half-power beam width (HPBW) of the telescope is 
$20.2''$ and $16.1''$ at 87 and 112 GHz, respectively, 
which correspond to 0.19 and 0.15 pc at the distance of W3 
\citep[1.95 kpc;][]{xu2006distance}, respectively. 
We observed two orthogonal polarization signals simultaneously 
by using the SIS mixer receiver (TZ1), 
whose system temperature ranged from 130 to 180 K. 
The backend was the autocorrelator SAM45. 
The frequency resolution and bandwidth are 488.28 kHz and 1600 MHz, respectively. 
The line intensity was calibrated by the chopper wheel method, 
and a typical calibration accuracy was 20\%. 
The antenna temperature is divided by the main beam efficiency of 
0.48 and 0.41 at 87 and 112 GHz, respectively, 
to obtain the main beam temperature $T_{\rm MB}$. 

We employed the On-The-Fly (OTF) mode for the mapping observation. 
The center position of the map is W3(OH): 
($\alpha_\mathrm{J2000.0}$, $\delta_\mathrm{J2000.0}$) = 
($2^\mathrm{h}27^\mathrm{m}4.0^\mathrm{s}$, $61^\circ52'24.0''$). 
The mapping area is a square of $16.0'\times16.0'$ centered at W3(OH), 
as shown in Figure \ref{optical}, 
which corresponds to the linear scale of 9.0 pc $\times$ 9.0 pc. 
The grid spacing for the OTF observation is $6''$. 
The total observation time was $\sim20$ hours. 
In addition to the OTF mapping, we conducted a short observation for a few minutes 
toward W3(OH) (the center position of the map) with the position-switching mode. 
Hereafter, we refer the spectrum of this position to as a ``hot core'' spectrum. 
For both the mapping and single-point observations, 
the reference position was taken as: 
($\alpha_\mathrm{J2000.0}$, $\delta_\mathrm{J2000.0}$) = 
($2^\mathrm{h}29^\mathrm{m}57.1^\mathrm{s}$, $62^\circ5'41.1''$). 
The telescope pointing was checked by observing 
the nearby SiO maser source (S Per) every hour, 
and the pointing accuracy was maintained to be better than $5''$. 

The observation data were first reduced with the NRO software 
\emph{NOSTAR} and \emph{NEWSTAR}, 
and then detailed analyses were carried out by using our own codes. 
Although we scanned two orthogonal direction 
($\alpha$- and $\delta$-direction) 
in the OTF mapping observation, 
we did not use the $\delta$-direction data for the analysis 
because of poor signal-to-noise ratios due to bad weather conditions. 
Since we convolved the image with the $30''$ beam for analysis, 
the scanning effects can be negligible. 

In the analysis, a baseline of the 5th-order polynomial 
was subtracted from each 0.2 GHz range, 
where the frequency resolution is 1 MHz. 
We identified the observed lines 
with the aid of the spectral line database CDMS 
\citep{muller2001cologne, muller2005cologne}. 
For the detailed analysis of each molecular line, 
a baseline of the 5th-order polynomial was subtracted 
in the velocity range from $-150$ to 50 km s$^{-1}$
with the velocity resolution of 2 km s$^{-1}$. 

\section{Results} \label{results}

\subsection{Hot-core-scale and molecular-cloud-scale spectra} \label{spectra}

\begin{figure*}
\begin{center}
\vspace{-40mm}
\includegraphics[width=\hsize, bb=0 0 720 432]{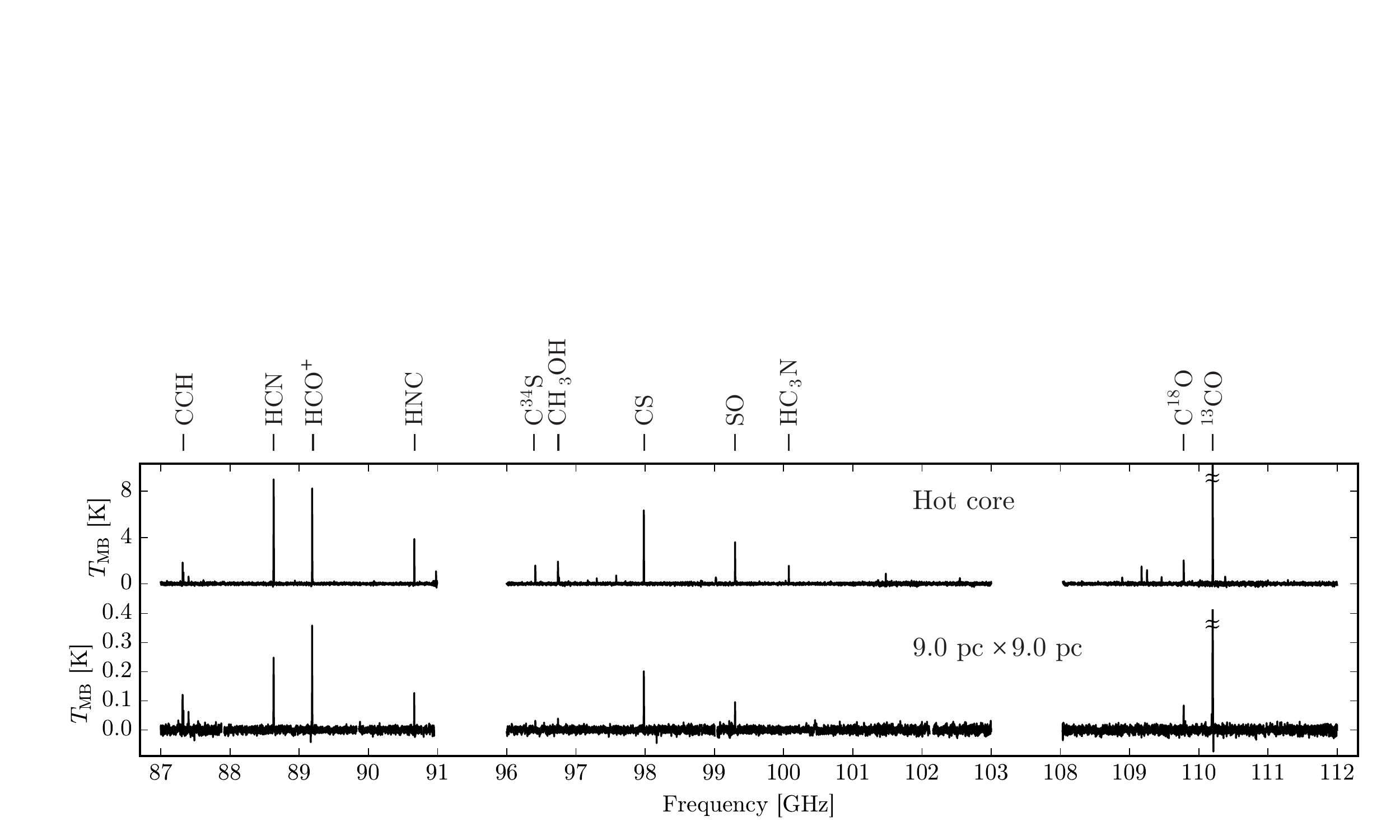}
\caption{\emph{Upper:} Spectrum of the W3(OH) hot core.
\emph{Lower:} The spectrum averaged over all the observed area (9.0 pc $\times$ 9.0 pc). 
The representative molecular lines are indicated on the top. 
\label{spectra_single-960}}
\end{center}
\end{figure*}

\begin{figure*}
\begin{center}
\includegraphics[width=\hsize, bb=0 0 720 268]{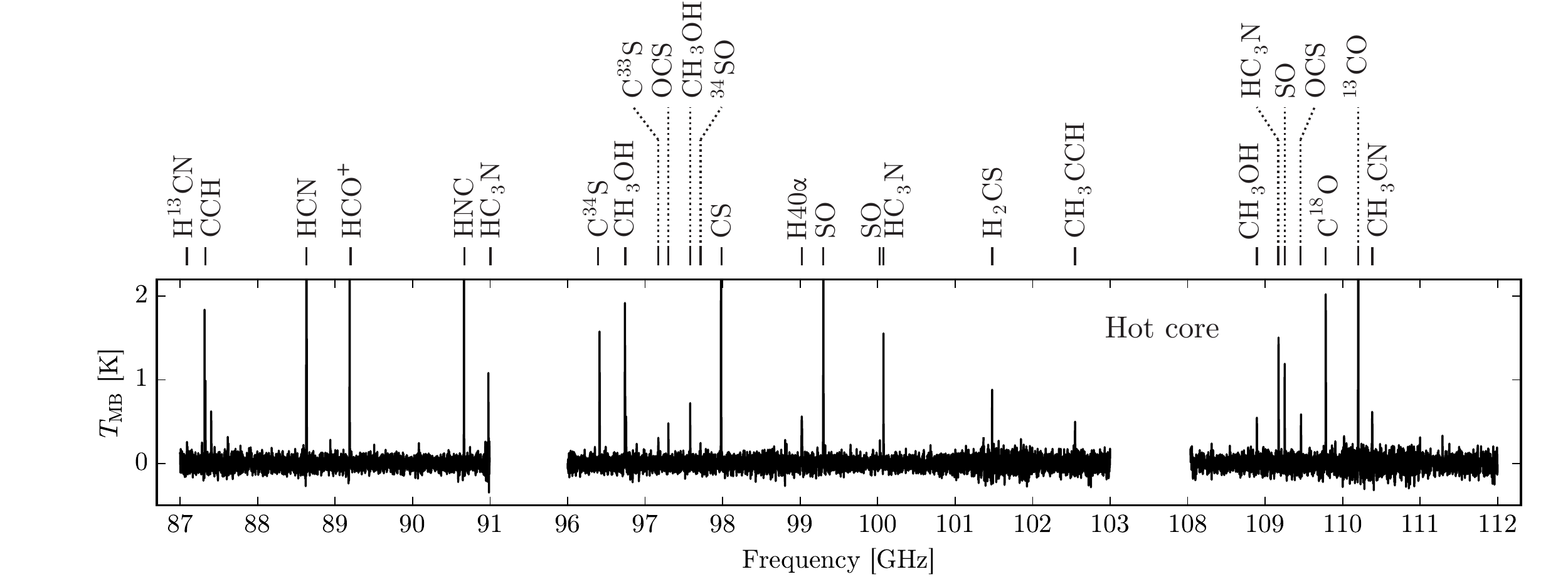}
\caption{Zoomed spectrum of the W3(OH) hot core. 
The detected species are indicated on the top.  
\label{spectra_single}}
\end{center}
\end{figure*}

\begin{figure*}
\begin{center}
\includegraphics[width=\hsize, bb=0 0 1080 576]{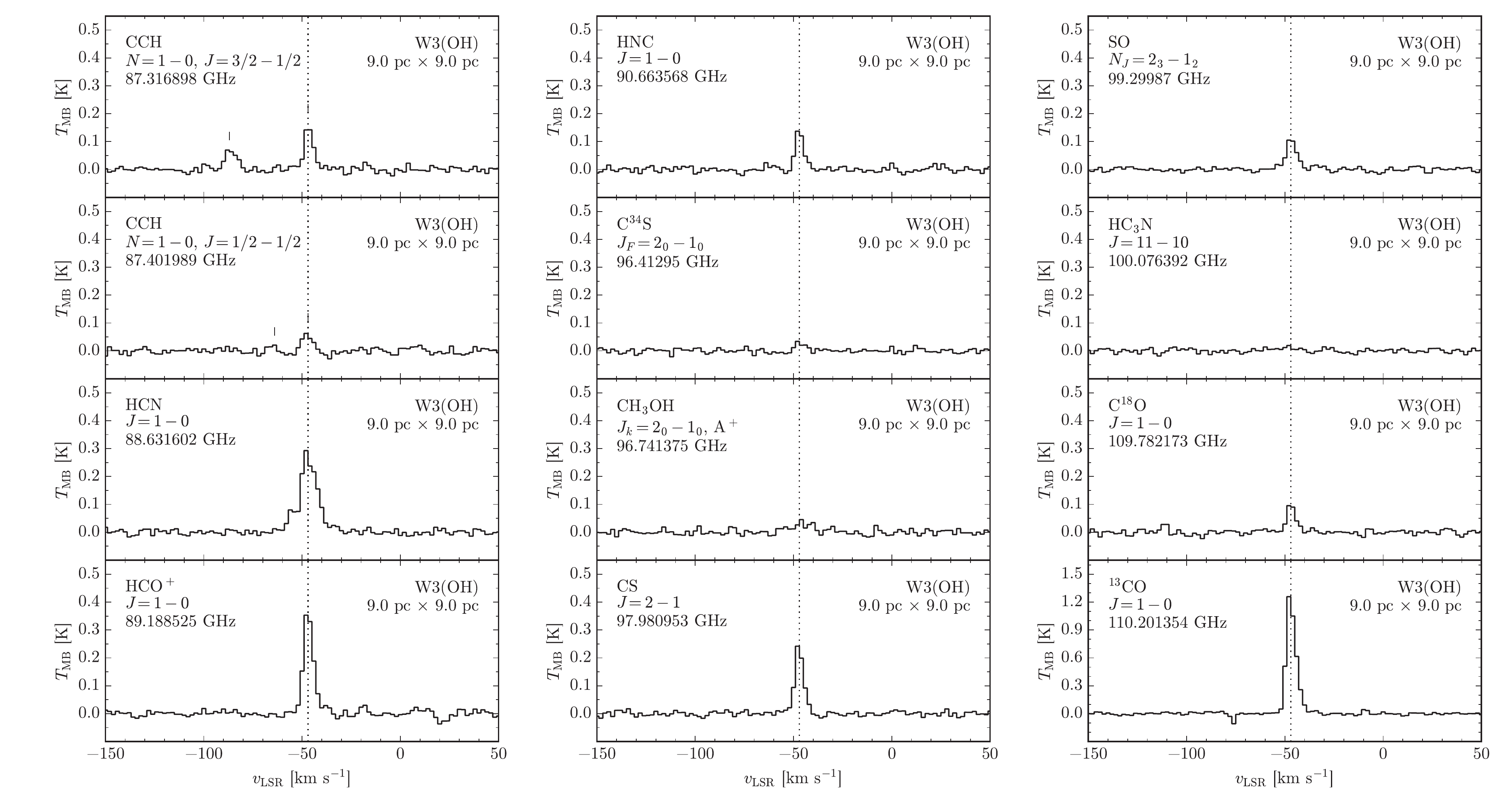}
\caption{Spectral line profiles of individual molecular transitions 
in the spectrum averaged over the whole area (9.0 pc $\times$ 9.0 pc). 
The dashed lines indicate the velocity of $-47$ km s$^{-1}$. 
Small vertical lines in the CCH panel represent the positions of the hyperfine components. 
\label{eachline960}}
\end{center}
\end{figure*}

Figure \ref{spectra_single-960} shows the ``hot core'' spectrum and 
the spectrum averaged over the full 9.0 pc $\times$ 9.0 pc area. 
The averaged spectrum is prepared 
by averaging the spectra of all the mapping grid points 
whose spacing is much smaller than the telescope beam. 
The lines of CCH, HCN, HCO$^+$, HNC, CS, SO, C$^{18}$O, and $^{13}$CO are 
strongly detected in the both spectra. 
In the ``hot core'' spectrum, the lines of C$^{34}$S, CH$_3$OH, and HC$_3$N are 
also detected with relatively high intensity. 
Furthermore, H$^{13}$CN, C$^{33}$S, OCS, $^{34}$SO, 
H$_2$CS, CH$_3$CCH, and CH$_3$CN are identified 
in spite of the short integration time, as shown in Figure \ref{spectra_single}. 
On the other hand, a fewer molecular lines are 
detected in the averaged spectrum. 
Even C$^{34}$S and CH$_3$OH are marginally detected, and HC$_3$N is absent. 
Individual line profiles 
in the spectrum averaged over the whole area 
are shown in Figure \ref{eachline960}. 
The line parameters obtained by the Gaussian fitting 
are summarized in Table \ref{lineparameters}. 

Generally, brightness temperatures of the averaged spectrum 
are fainter than those of the ``hot core'' spectrum. 
However, intensity ratios of molecular lines between the ``hot core'' spectrum 
and the averaged spectrum are different from species to species. 
Using HCO$^+$ as a reference, 
CCH is found to be relatively strong in the averaged spectrum, 
while the other species such as HCN, HNC, CS, and SO are fainter. 
This fact may reflect the different distributions of these emissions. 
We will discuss this point more closely in Section \ref{subregion}. 
OCS, H$_2$CS, CH$_3$CCH, and CH$_3$CN are detected 
only in the ``hot core'' spectrum. 
These species would be mainly distributed just around the star-forming region. 

\begin{deluxetable*}{lrlrrrrr}
\tabletypesize{\scriptsize}
\tablecaption{Line parameters of the spectrum averaged over all the observed area 
(9.0 pc $\times$ 9.0 pc). \label{lineparameters}}
\tablewidth{0pt}
\tablehead{
\colhead{Molecule} & \colhead{Frequency} 
& \colhead{Transition} & \colhead{$T_{\rm MB}$ Peak} 
& \colhead{$v_{\rm LSR}$} & \colhead{$\Delta v$}
& \colhead{$\int T_{\rm MB}dv$} \\
\colhead{} & \colhead{(GHz)} 
& \colhead{} & \colhead{(mK)} & \colhead{(km s$^{-1}$)} 
& \colhead{(km s$^{-1}$)} & \colhead{(K km s$^{-1}$)}}
\startdata
CCH & 87.284105 & $N=1-0$, $J=3/2-1/2$, $F=1-1$ &
&
&
&
$<0.1$ \\
CCH & 87.316898 & $N=1-0$, $J=3/2-1/2$, $F=2-1$ &
$ 0.157 \pm 0.013 $ &
$ -46.6 \pm 0.2 $ &
$ 5.2 \pm 0.5 $ &
$ 0.95 \pm 0.03 $ \\
CCH & 87.328585 & $N=1-0$, $J=3/2-1/2$, $F=1-0$ &
$ 0.072 \pm 0.018 $ &
$ -46.1 \pm 0.9 $ &
$   7 \pm   2 $ &
$ 0.51 \pm 0.03 $ \\
CCH & 87.401989 & $N=1-0$, $J=1/2-1/2$, $F=1-1$ &
$ 0.062 \pm 0.008 $ &
$ -47.6 \pm 0.4 $ &
$ 5.9 \pm 0.9 $ &
$ 0.30 \pm 0.03 $ \\
CCH & 87.407165 & $N=1-0$, $J=1/2-1/2$, $F=0-1$ &
$ 0.024 \pm 0.013 $ &
$ -47.7 \pm 1.2 $ &
$   5 \pm   3 $ &
$ 0.05 \pm 0.03 $ \\
CCH & 87.446470 & $N=1-0$, $J=1/2-1/2$, $F=1-0$ &
&
&
&
$<0.1$ \\
HCN & 88.631602 & $J=1-0$ &
$ 0.263 \pm 0.008 $ &
$ -46.71 \pm 0.15 $ &
$ 10.6 \pm 0.4 $ &
$ 3.16 \pm 0.03 $ \\
HCO$^+$ & 89.188525 & $J=1-0$ &
$ 0.368 \pm 0.010 $ &
$ -46.85 \pm 0.08 $ &
$ 5.83 \pm 0.18 $ &
$ 2.37 \pm 0.02 $ \\
HNC & 90.663568 & $J=1-0$ &
$ 0.146 \pm 0.009 $ &
$ -47.05 \pm 0.14 $ &
$ 4.7 \pm 0.3 $ &
$ 0.71 \pm 0.02 $ \\
C$^{34}$S & 96.412950 & $J_F=2_0-1_0$ &
$ 0.030 \pm 0.006 $ &
$ -46.6 \pm 0.6 $ &
$ 5.8 \pm 1.4 $ &
$ 0.148 \pm 0.019 $ \\
CH$_3$OH & 96.741375 & $J_K=2_0-1_0$, A$^+$ &
$ 0.030 \pm 0.005 $ &
$ -45.5 \pm 1.1 $ &
$  14 \pm   3 $ &
$ 0.42 \pm 0.02 $ \\
CS & 97.980953 & $J=2-1$ &
$ 0.244 \pm 0.007 $ &
$ -47.15 \pm 0.08 $ &
$ 5.19 \pm 0.18 $ &
$ 1.38 \pm 0.02 $ \\
SO & 99.299870 & $N_J=2_3-1_2$ &
$ 0.108 \pm 0.006 $ &
$ -46.71 \pm 0.16 $ &
$ 6.3 \pm 0.4 $ &
$ 0.751 \pm 0.019 $ \\
HC$_3$N & 100.076392 & $J=11-10$ &
&
&
&
$< 0.07$ \\
C$^{18}$O & 109.782173 & $J=1-0$ &
$ 0.102 \pm 0.008 $ &
$ -46.86 \pm 0.19 $ &
$ 5.1 \pm 0.4 $ &
$ 0.58 \pm 0.02 $ \\
$^{13}$CO & 110.201354 & $J=1-0$ &
$ 1.25 \pm 0.02 $ &
$ -46.93 \pm 0.05 $ &
$ 5.70 \pm 0.13 $ &
$ 7.74 \pm 0.02 $ 
\enddata
\tablecomments{The errors are 1$\sigma$. The upper limits are 3$\sigma$. 
The calibration error ($\sim20$\%) is not included. 
A peak temperature ($T_\mathrm{MB}$ Peak), 
a peak velocity ($v_\mathrm{LSR}$), 
and a line width ($\Delta v$) are obtained by gaussian fit 
to the spectrum averaged over the whole observed area 
(shown in Figure \ref{eachline960}). 
Integrated intensities ($\int T_{\rm MB}dv$) are calculated 
in the velocity range from $-57$ to $-37$ km s$^{-1}$ except for HCN. 
For HCN, the velocity range is from $-67$ to $-27$ km s$^{-1}$ 
to cover all the hyperfine components. }
\end{deluxetable*}

\subsection{Distribution of key molecular species} \label{distribution}

\begin{figure*}
\begin{center}
\includegraphics[width=0.9\hsize, bb=0 0 553 669]{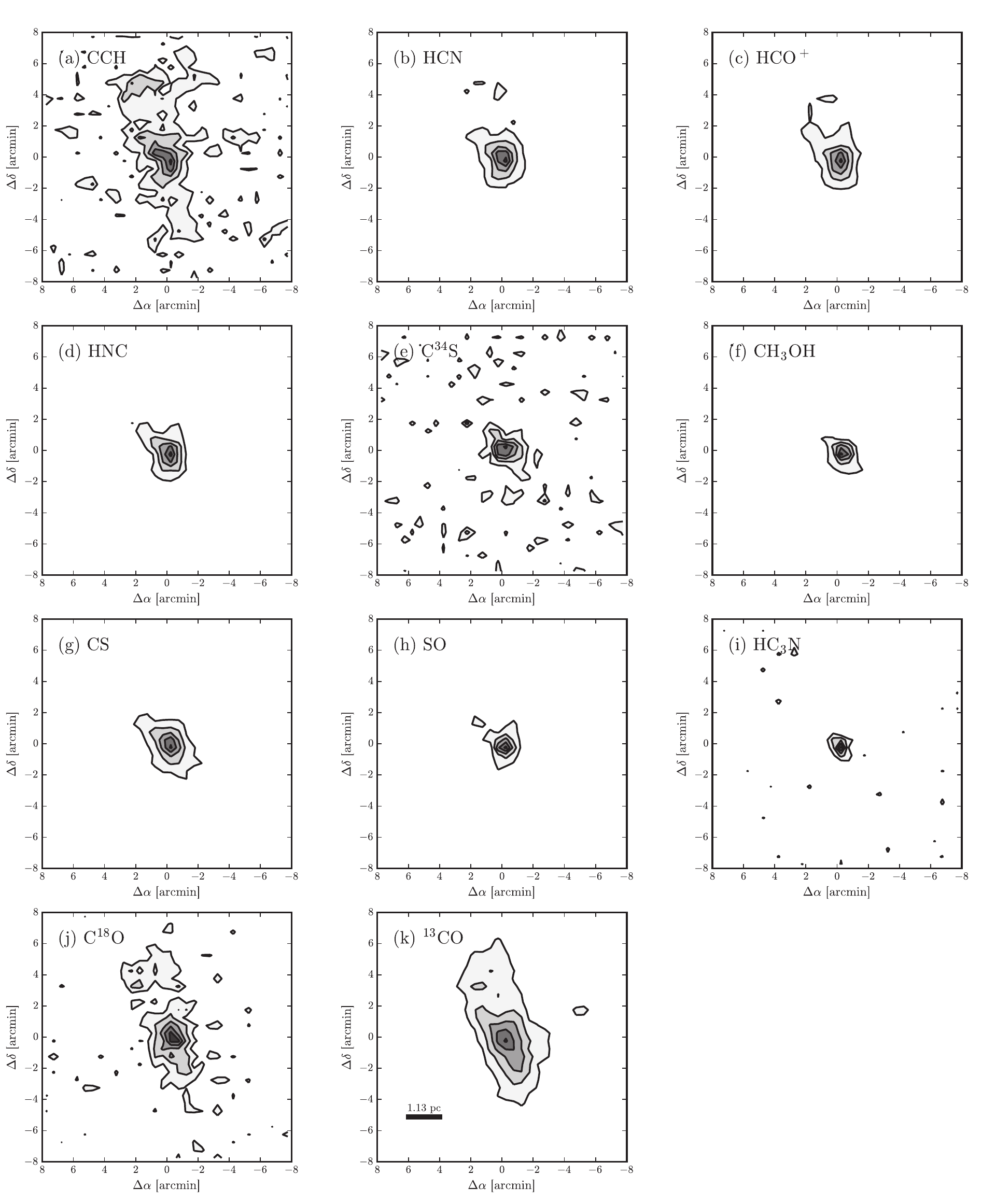}
\caption{Integrated intensity maps of the bright emission lines. 
The coordinates ($\Delta\alpha$, $\Delta\delta$) are the offsets from W3(OH) 
($\alpha_\mathrm{J2000.0}=2^\mathrm{h}27^\mathrm{m}4.0^\mathrm{s}$, 
$\delta_\mathrm{J2000.0}=61^\circ52'24.0''$). 
The images are convolved to the $30''$ resolution. 
The velocity range of integration is from $-57$ to $-37$ km s$^{-1}$ 
except for HCN. 
For HCN, the velocity interval is from $-67$ to $-27$ km s$^{-1}$ 
to cover all the hyperfine components. 
The contour levels are (20\%, 40\%, 60\%, 80\%, 100\%) of 
(a) 10 K km s$^{-1}$, (b) 70 K km s$^{-1}$, (c) 50 K km s$^{-1}$, 
(d) 25 K km s$^{-1}$, (e) 6 K km s$^{-1}$, (f) 18 K km s$^{-1}$, 
(g) 38 K km s$^{-1}$, (h) 25 K km s$^{-1}$, (i) 10 K km s$^{-1}$, 
(j) 10 K km s$^{-1}$, and (k) 90 K km s$^{-1}$. 
\label{map}}
\end{center}
\end{figure*}

In order to characterize the chemical composition 
averaged over the 9.0 pc $\times$ 9.0 pc area, 
it is necessary to investigate which molecular emission comes 
from which part of the molecular cloud. 
We select the 11 strongest lines, CCH ($N=1-0$), 
HCN ($J=1-0$), HCO$^+$ ($J=1-0$), HNC ($J=1-0$), 
C$^{34}$S ($J=2-1$), CH$_3$OH ($J_K=2_K-1_K$), 
CS ($J=2-1$), SO ($J_K=3_2-2_1$), HC$_3$N ($J=11-10$), 
C$^{18}$O ($J=1-0$), and $^{13}$CO ($J=1-0$), 
and prepare their integrated intensity maps with the $30''$ grid,  
as shown in Figure \ref{map}. 
This grid size corresponds to the spatial scale of 0.28 pc 
at the distance of W3 (1.95 kpc). 
The velocity range of integration is from $-57$ to $-37$ km s$^{-1}$ 
except for HCN. 
For HCN, the velocity range is from $-67$ to $-27$ km s$^{-1}$ 
in order to cover all the hyperfine components. 
It should be noted that the velocity difference across the observed area 
is 4 km s $^{-1}$ at most, which is smaller than 
the range of integration mentioned above. 

The distribution of $^{13}$CO extends to the north-south direction, 
which is consistent with the $^{13}$CO map 
previously reported by \citet{sakai2006atomic}. 
The distribution of C$^{18}$O is also consistent 
with the map by \citet{sakai2006atomic}, 
in spite of the rather poor signal-to-noise ratio of our C$^{18}$O data. 
All the distributions including these of $^{13}$CO and C$^{18}$O 
have a peak at the center position. 
However, the degree of concentration to the center 
differs from species to species. 
To evaluate the degree of concentration, 
we apply a two-dimensional Gaussian fit to integrated intensities. 
The parameters are summarized in Table \ref{2Dgauss}. 
We can arbitrarily classify the distributions into the following three types; 
(i) species concentrated just around the center, 
such as CH$_3$OH, SO, and HC$_3$N, 
(ii) species loosely concentrated around the center, 
such as HCN, HCO$^+$, HNC, and CS, 
and (iii) species extended widely, such as CCH, C$^{18}$O, and $^{13}$CO. 

Such a difference of the emitting region among molecular species 
is caused by excitation effects and chemical abundance variations. 
Although the critical density of molecular line is 
an important factor to the difference, it cannot explain everything. 
Table \ref{critical} summarizes the critical density 
and the upper state energy of each spectral line. 
The HCN ($J=1-0$) line has the highest critical density 
among the lines listed in Table \ref{critical}. 
However, it is not exactly true that the HCN line 
only reflects high density regions in the molecular-cloud-scale spectrum. 
In fact, the HCN emission is more extended than the emission 
of other molecular lines with lower critical densities. 
The HCN molecules can sub-thermally be excited 
even in a region below the critical density, 
and hence the HCN emission can come from such less dense regions 
in the observation beam to some extent, as far as HCN exists. 
This is indeed revealed for the Orion A molecular cloud 
by \citet{kauffmann2017molecular}. 
Since the less dense part has a larger area of molecular clouds, 
its contribution to the molecular-cloud-scale spectrum 
could overwhelm the contribution from the dense part. 
Needless to say, whether this situation happens 
also depends on the molecular abundance in each part. 
Thus, molecular-cloud-scale spectrum is 
also sensitive to chemical abundance variation, 
which is caused by various factors such as the H$_2$ density, 
the gas kinetic temperature, and the UV radiation field. 
This is the reason why we observationally 
study the molecular-cloud-scale spectrum 
by conducting mapping observations toward Galactic sources. 
We will discuss on this issue in Section \ref{correlation}. 

\begin{deluxetable*}{lrlccc}
\tabletypesize{\scriptsize}
\tablecaption{Spatial extent of key molecular species. 
\label{2Dgauss}}
\tablewidth{0pt}
\tablehead{
\colhead{Molecule} & \colhead{Frequency} & \colhead{Transition} & 
\colhead{Major axis} & \colhead{Minor axis} & \colhead{Position angle}\\
\colhead{} & \colhead{(GHz)} & \colhead{} & 
\colhead{(arcmin)} & \colhead{(arcmin)} & \colhead{(deg)}}
\startdata
CCH & 87.316898 & $N=1-0$, $J=5/2-3/2$, $F=2-1$ &
4.2 & 2.0 & 30 \\
$^{13}$CO & 110.201354 & $J=1-0$ &
3.2 & 1.3 & 30 \\
C$^{18}$O & 109.782173 & $J=1-0$ &
2.8 & 1.1 & 30 \\
\hline
HCO$^+$ & 89.188525 & $J=1-0$ &
2.2 & 1.3 & 30 \\
HCN & 88.631602 & $J=1-0$ &
1.9 & 1.2 & 30 \\
CS & 97.980953 & $J=2-1$ &
1.5 & 1.0 & 30 \\
HNC & 90.663568 & $J=1-0$ &
1.4 & 0.9 & 30 \\
\hline
SO & 99.299870 & $N_J=2_3-1_2$ &
1.2 & 0.7 & 30 \\
C$^{34}$S & 96.412950 & $J_F=2_0-1_0$ &
1.0 & 0.7 & 78 \\
CH$_3$OH & 96.741375 & $J_K=2_0-1_0$, A$^+$ &
1.0 & 0.6 & 33 \\
HC$_3$N & 100.076392 & $J=11-10$ &
0.7 & 0.4 & 55 \\
\enddata
\tablecomments{The FWHM values of major axis and minor axis 
are derived by the two-dimensional Gaussian fit. 
The position angle is measured from north through east. }
\end{deluxetable*}

\begin{deluxetable*}{lrlccc}
\tabletypesize{\scriptsize}
\tablecaption{The critical density and upper state energy of each spectral line. 
\label{critical}}
\tablewidth{0pt}
\tablehead{
\colhead{Molecule} & \colhead{Frequency} & \colhead{Transition} & 
\colhead{$E_u$} & \colhead{$n_{crit}$ at 10 K} & \colhead{$n_{crit}$ at 100 K}\\
\colhead{} & \colhead{(GHz)} & \colhead{} & 
\colhead{(K)} & \colhead{(cm$^{-3}$)} & \colhead{(cm$^{-3}$)}}
\startdata
CCH & 87.316898 & $N=1-0$, $J=5/2-3/2$, $F=2-1$ &
4.2 & $1.21\times10^5$ & $1.94\times10^5$ \\
HCN & 88.631602 & $J=1-0$ &
4.3 & $1.01\times10^6$ & $2.66\times10^6$ \\
HCO$^+$ & 89.188525 & $J=1-0$ &
4.3 & $1.64\times10^5$ & $2.36\times10^5$ \\
HNC & 90.663568 & $J=1-0$ &
4.4 & $2.77\times10^5$ & $4.12\times10^5$ \\
CH$_3$OH & 96.741375 & $J_K=2_0-1_0$, A$^+$ &
7.0 & $3.10\times10^4$ & $3.83\times10^4$ \\
CS & 97.980953 & $J=2-1$ &
7.1 & $3.32\times10^5$ & $3.84\times10^5$ \\
SO & 99.299870 & $N_J=2_3-1_2$ &
9.2 & $\cdots$ & $2.95\times10^5$ \\
HC$_3$N & 100.076392 & $J=11-10$ &
28.8 & $8.83\times10^5$ & $\cdots$ \\
H$_2$CS & 101.477810 & $J_{K_p,K_o}=3_{1,3}-2_{1,2}$ &
22.9 & $1.59\times10^5$ & $1.77\times10^5$\\
OCS & 109.463063 & $J=9-8$ &
26.3 & $5.43\times10^4$ & $4.56\times10^4$\\
C$^{18}$O & 109.782173 & $J=1-0$ &
5.3 & $1.90\times10^3$ & $1.80\times10^3$ \\
$^{13}$CO & 110.201354 & $J=1-0$ &
5.3 & $1.91\times10^3$ & $1.80\times10^3$\\
CH$_3$CN & 110.383505 & $J_K=6_0-5_0$ &
18.5 & $\cdots$ & $5.50\times10^5$
\enddata
\tablecomments{The critical density $n_{crit}$ is derived by $A_{ij}/C_{ij}$, where 
$A_{ij}$ is the Einstein A-coefficient 
and $C_{ij}$ is the collisional rate coefficient. 
These are adopted from the LAMDA database \citep{vandertak2007computer} 
and the original papers: 
CCH: \citet{spielfiedel2012fine}; 
HCN and HNC: \citet{dumouchel2010rotational}; 
HCO$^+$: \citet{flower1999rotational}; 
CH$_3$OH: \citet{rabli2010rotational}; 
CS and SO: \citet{lique2006rotational}; 
HC$_3$N and OCS: \citet{green1978collisional}; 
H$_2$CS: \citet{wiesenfeld2013rotational}; 
C$^{18}$O and $^{13}$CO: \citet{yang2010rotational}; 
CH$_3$CN: \citet{green1986collisional}. 
For C$^{34}$S, the collisional rates are not available in the LAMDA. }
\end{deluxetable*}

\section{Discussion} \label{discussion}

\subsection{Correlation between the integrated intensities of molecules 
and $^{13}$CO} \label{correlation}

\begin{figure*}
\begin{center}
\includegraphics[width=0.9\hsize, bb=0 0 550 709]{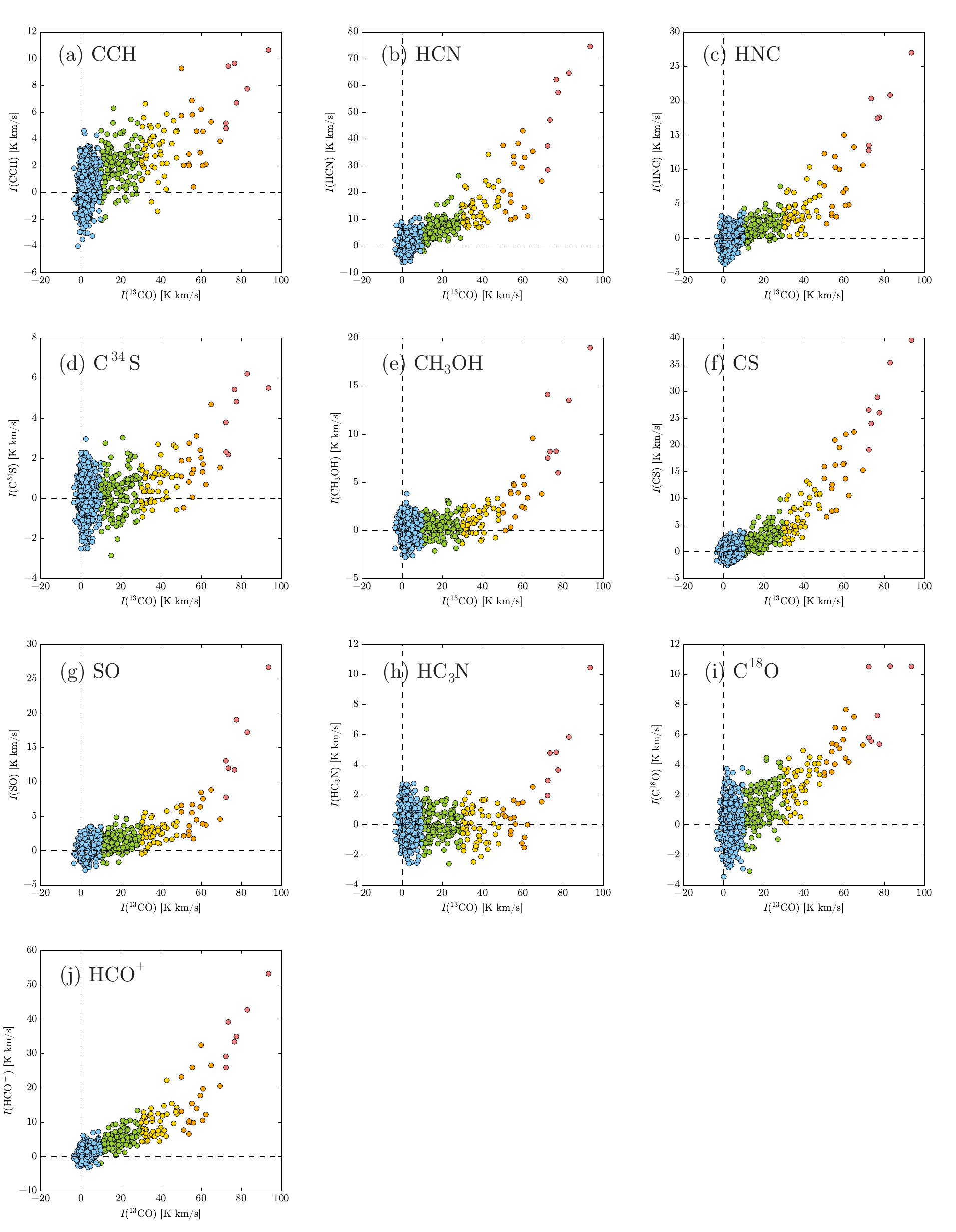}
\caption{Correlation diagrams of integrated intensities of various molecules 
against that of $^{13}$CO. 
The velocity range of integration is 
from $-57$ to $-37$ km s$^{-1}$ except for HCN. 
For HCN, the velocity interval is from $-67$ to $-27$ km s$^{-1}$ 
to cover all the hyperfine components. 
The different colors indicate 5 subregions classified by 
the $^{13}$CO integrated intensity: 
\emph{red}: subregion A ($>70$ K km s$^{-1}$), 
\emph{orange}: subregion B ($50-70$ K km s$^{-1}$), 
\emph{yellow}: subregion C ($30-50$ K km s$^{-1}$), 
\emph{light green}: subregion D ($10-30$ K km s$^{-1}$), 
and \emph{light blue}: subregion E ($<10$ K km s$^{-1}$). 
\label{integint_correlation}}
\end{center}
\end{figure*}

As mentioned in Section \ref{distribution}, 
the distribution is different among the observed molecular species. 
In order to characterize the difference, we investigate correlations 
between the integrated intensities of $^{13}$CO and those of various molecules. 
Here, we employ $^{13}$CO as a reference molecule, 
because the $^{13}$CO integrated intensity 
would roughly represent the line-of-sight column density of H$_2$. 
Figure \ref{integint_correlation} shows the correlation diagrams. 
Each data point represents the integrated intensity 
averaged over a $30''\times30''$ (0.28 pc $\times$ 0.28 pc) area. 
Consequently, the number of the data for 
the whole mapped area ($16'\times16'$) is 1024 in total. 
Note that the typical error of the integrated intensity at 
each data point is 2 K km s$^{-1}$ (3$\sigma$). 

Roughly speaking, all diagrams show positive correlation. 
However, the dependence on the $^{13}$CO integrated intensity 
is different from molecule to molecule, 
as expected from the variation of the molecular distributions. 
As shown in Figures \ref{integint_correlation}e and \ref{integint_correlation}h, 
the integrated intensities of CH$_3$OH and HC$_3$N 
seem to have an appearance threshold at the $^{13}$CO integrated intensity 
of about 40 K km s$^{-1}$ and 60 K km s$^{-1}$, respectively. 
The integrated intensities of C$^{34}$S and SO gradually increase 
with the increasing $^{13}$CO integrated intensity 
at $I(^{13}\mathrm{CO})<60$ K km s$^{-1}$, 
but the slope becomes steeper at a higher $^{13}$CO intensity 
(Figures \ref{integint_correlation}d and \ref{integint_correlation}g).
These behaviors are consistent with the fact that 
the above four molecular species, CH$_3$OH, HC$_3$N, C$^{34}$S, 
and SO, are mainly concentrated around the W3(OH) hot core (Figure \ref{map}). 

The integrated intensities of HCN, HCO$^+$, HNC, and CS start to rise 
with that of $^{13}$CO more steeply than those of C$^{34}$S and SO 
for $I(^{13}\mathrm{CO})$ less than 40 K km s$^{-1}$ 
(Figures \ref{integint_correlation}b, \ref{integint_correlation}j, 
\ref{integint_correlation}c, and \ref{integint_correlation}f). 
However, their dependence on the $^{13}$CO integrated intensity is not linear, 
but seems to have a knee point at the $^{13}$CO integrated intensity 
of about $40-60$ K km s$^{-1}$, above which the dependence becomes steeper. 
Similarity of these four species in their dependences on 
the $^{13}$CO integrated intensity indicates their similar spatial distribution, 
as shown in Figure \ref{map}. 

While the integrated intensities of most molecular species 
show non-linear dependences on the $^{13}$CO integrated intensity, 
the integrated intensities of C$^{18}$O and CCH 
increase almost linearly as the increasing $^{13}$CO integrated intensity 
(Figures \ref{integint_correlation}i and \ref{integint_correlation}a). 
The linear correlation between C$^{18}$O and $^{13}$CO means that 
the $^{13}$CO line is not optically very thick. 
In fact, the $^{13}$CO/C$^{18}$O ratio is approximately 10, 
which is consistent with the $^{13}$C/$^{18}$O ratio 
in the local interstellar medium \citep[$11\pm2$;][]{lucas1998interstellar}. 
Hence, it is confirmed that the rapid increase of the integrated intensities 
of various molecules for the $^{13}$CO integrated intensity higher than 
$40-60$ K km s$^{-1}$ does not mean the saturation of the $^{13}$CO line. 
CCH also shows a linear correlation with $^{13}$CO, 
although the signal-to-noise ratio is not very good. 
This result suggests that CCH is widely distributed 
over the molecular cloud as $^{13}$CO. 

\begin{figure*}
\begin{center}
\includegraphics[width=0.9\hsize, bb=0 0 550 709]{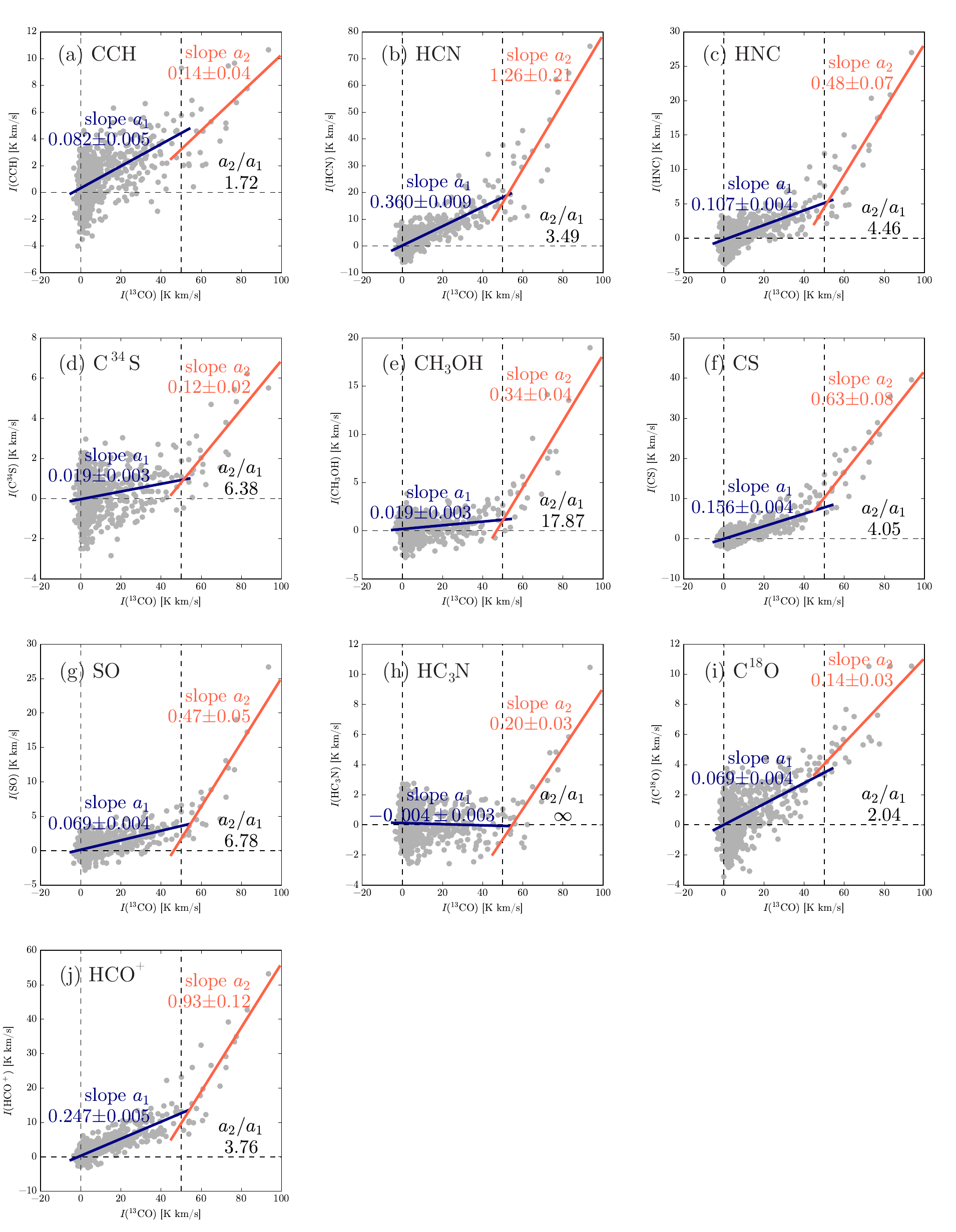}
\caption{Same as the Figure \ref{integint_correlation}, 
but with linear function fittings. 
Navy line is a linear function fitting for the data 
with the $^{13}$CO integrated intensities of $<50$ K km s$^{-1}$, 
whereas orange-red line is that of $>50$ K km s$^{-1}$. 
The slopes $a_1$ and $a_2$ correspond to the fits for 
$<50$ K km s$^{-1}$ and $>50$ K km s$^{-1}$, respectively. 
The $a_2/a_1$ ratio roughly indicates how much the slope is 
increased across the knee point ($\sim50$ K km s$^{-1}$): 
a larger $a_2/a_1$ ratio means a larger change in the slope. 
\label{integint_correlation_fit}}
\end{center}
\end{figure*}

To quantitatively reveal the different behavior 
of the different species in Figure \ref{integint_correlation}, 
we conduct a phenomenological fit 
to the correlation diagrams (Figure \ref{integint_correlation_fit}). 
Considering that most molecular species have a knee point around 
the $^{13}$CO integrated intensity of 50 K km s$^{-1}$, 
we employ two different linear functions for 
the ranges lower and higher than 50 K km s$^{-1}$. 
The slopes $a_1$ and $a_2$ shown in Figure \ref{integint_correlation_fit} 
correspond to the fits for the lower and higher ranges, respectively. 
Hence, the $a_2/a_1$ ratio roughly indicates how much the slope is 
increased from the lower range to the higher range: 
a larger $a_2/a_1$ ratio means a larger change in the slope. 
While the $a_2/a_1$ ratios of HCN, HCO$^+$, HNC, and CS are around 4, 
that of CH$_3$OH, HC$_3$N, C$^{34}$S, and SO are higher than 6. 
This result quantitatively shows a steeper increase of the intensities 
of CH$_3$OH, HC$_3$N, C$^{34}$S, and SO across the knee point. 
Note that this linear-function fitting is not based on any theoretical models, 
and the criteria of 50 K km s$^{-1}$ is arbitrary. 

\begin{figure*}
\begin{center}
\includegraphics[width=0.9\hsize, bb=0 0 550 533]{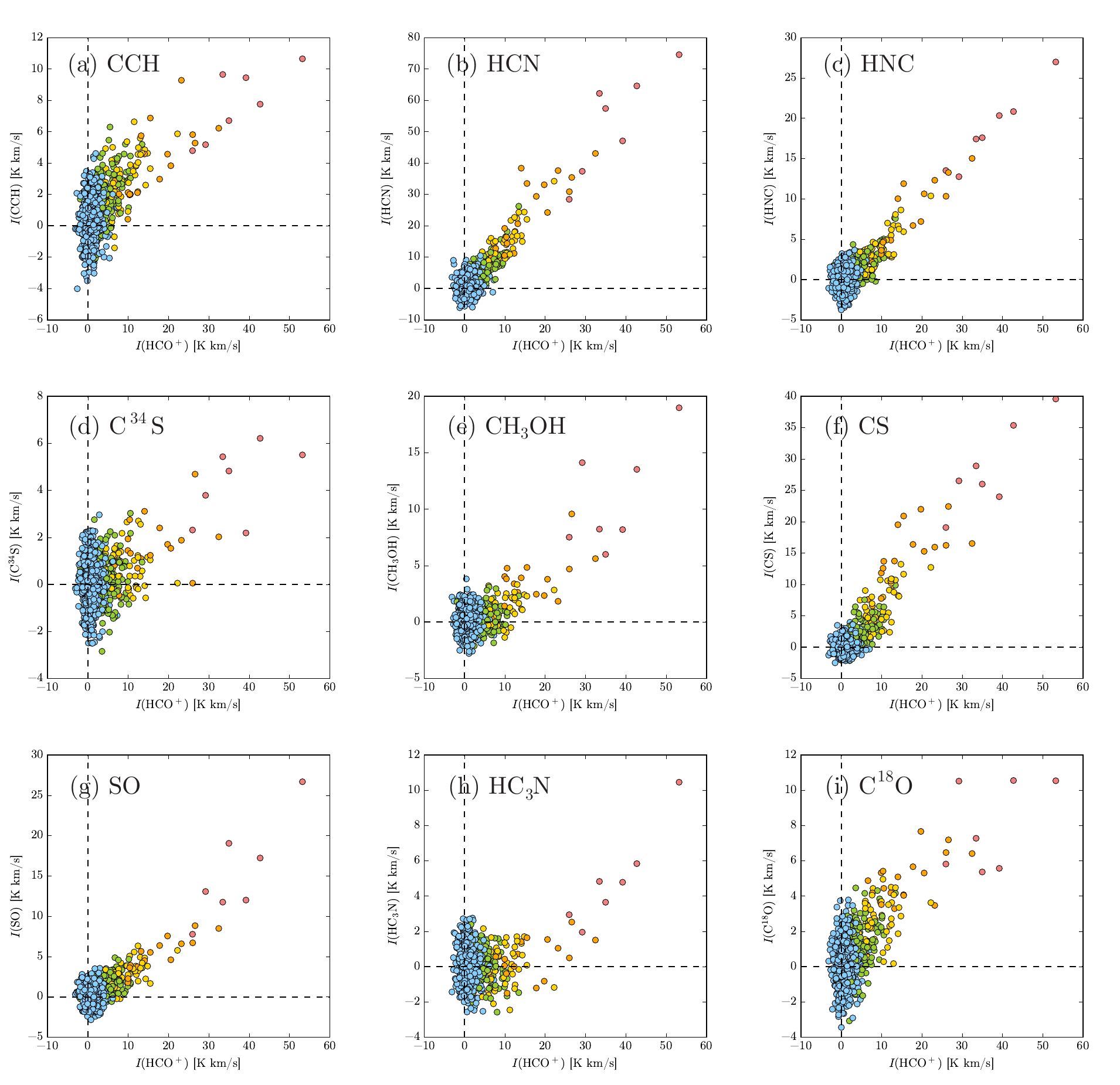}
\caption{Correlation diagrams of integrated intensities of various molecules 
against that of HCO$^+$. 
The colors correspond to the subregions 
classified by the integrated intensity of $^{13}$CO. 
The velocity range of integration is from $-57$ to $-37$ km s$^{-1}$ 
except for HCN. 
For HCN, the velocity interval is from $-67$ to $-27$ km s$^{-1}$ 
to cover all the hyperfine components. 
The different colors indicate 5 subregions classified by 
the $^{13}$CO integrated intensity: 
\emph{red}: subregion A ($>70$ K km s$^{-1}$); 
\emph{orange}: subregion B ($50-70$ K km s$^{-1}$); 
\emph{yellow}: subregion C ($30-50$ K km s$^{-1}$); 
\emph{light green}: subregion D ($10-30$ K km s$^{-1}$); 
and \emph{light blue}: subregion E ($<10$ K km s$^{-1}$). 
\label{integint_correlation_HCOp}}
\end{center}
\end{figure*}

For further investigations, similar correlation diagrams 
are prepared against the HCO$^+$ integrated intensity 
instead of the $^{13}$CO integrated intensity (Figure \ref{integint_correlation_HCOp}). 
The integrated intensities of HCN, HNC, and CS are linearly correlated well, 
where the correlation coefficient is higher than 0.8 
(Figures \ref{integint_correlation_HCOp}b, \ref{integint_correlation_HCOp}c, 
and \ref{integint_correlation_HCOp}f). 
This again verifies similar distributions of HCO$^+$, HCN, HNC, and CS. 
The integrated intensity of CH$_3$OH, SO, and HC$_3$N seem to have a knee point 
at the HCO$^+$ integrated intensity of $20-30$ K km s$^{-1}$, 
as seen as the appearance threshold in the correlation diagram with $^{13}$CO 
(Figures \ref{integint_correlation_HCOp}e, \ref{integint_correlation_HCOp}g, 
and \ref{integint_correlation_HCOp}h). 
A similar trend is marginally seen for C$^{34}$S 
(Figure \ref{integint_correlation_HCOp}d) 
in spite of the poor signal-to-noise ratio. 
On the other hand, the CCH and C$^{18}$O emissions are brightly observed 
even for the low HCO$^+$ integrated intensity range 
(Figure \ref{integint_correlation_HCOp}a and \ref{integint_correlation_HCOp}i). 
The CCH and C$^{18}$O emission at the low HCO$^+$ integrated intensity 
seem significant, because their noise levels are 
$2$ K km s$^{-1}$ (3$\sigma$). 

The emission from the extended gas 
with the low $^{13}$CO integrated intensity is generally faint, 
but the spatial area of such regions is much larger than that of dense cores. 
When the extended component is accumulated, 
it cannot be ignored in the averaged spectrum. 
In the following section, we examine the contributions of extended gas components 
and dense cores by classifying the observed area into 5 subregions 
according to the $^{13}$CO integrated intensity. 

\subsection{5 subregions and their characteristics} \label{subregion}

\begin{figure*}
\begin{center}
\includegraphics[width=0.67\hsize, bb=0 0 360 180]{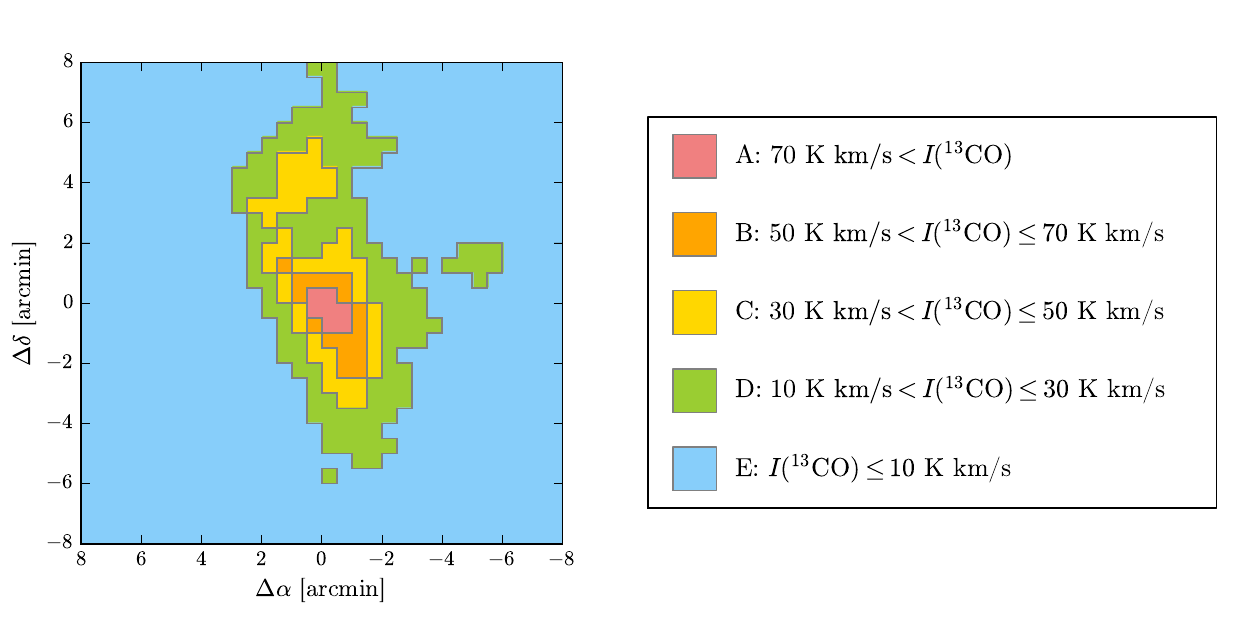}
\caption{5 subregions classified by the $^{13}$CO integrated intensity. 
The coordinates ($\Delta\alpha$, $\Delta\delta$) are 
the same as Figure \ref{map}. 
\label{map_region}}
\end{center}
\end{figure*}

What does the spectrum averaged over the molecular cloud represent? 
To address this question, we classify the full mapped region into 5 subregions 
according to the $^{13}$CO integrated intensity $I(^{13}\mathrm{CO})$. 
A criteria for the classification is arbitrary, as summarized in Table \ref{criteria}. 
The classification point is set every 20 K km s$^{-1}$, 
considering the highest $I(^{13}\mathrm{CO})$ of 93.5 K km s$^{-1}$. 
The 5 subregions (A$-$E) are shown in Figure \ref{map_region}. 
The subregion A is just around the W3(OH) hot core, 
while subregions B, C, and D surround it, extending along north-south direction. 
The subregion E is the other remaining part. 

\begin{figure*}
\begin{center}
\includegraphics[width=\hsize, bb=0 0 720 484]{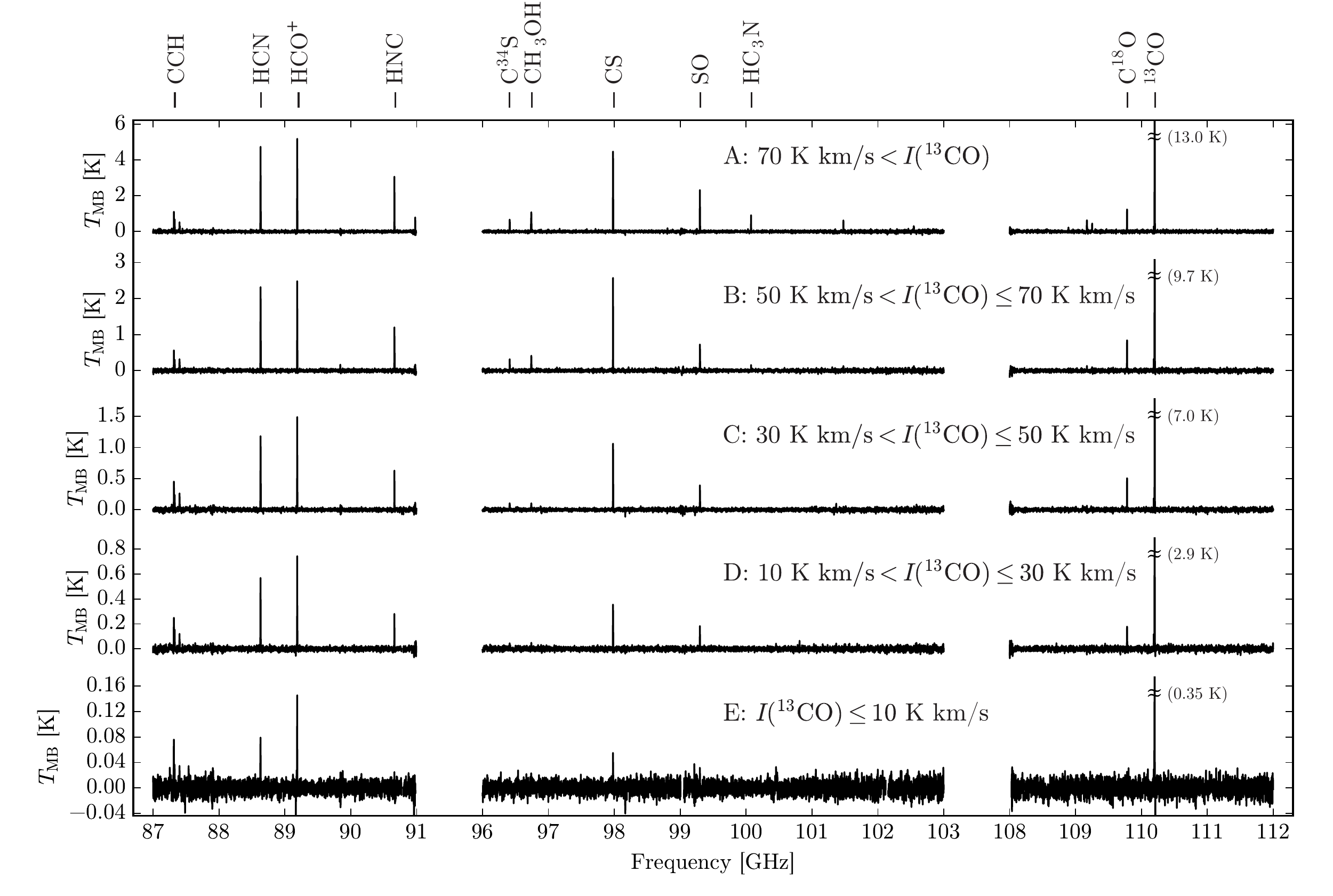}
\caption{Spectra of subregions A, B, C, D, and E. 
The apparent sizes of the HCO$^+$ line intensity are 
presented to be identical among the panels. 
Note that the temperature scale of the spectrum 
is different from subregion to subregion. 
\label{spectra_region}}
\end{center}
\end{figure*}

Then, we prepare a spatially averaged spectrum of each subregion, 
as shown in Figure \ref{spectra_region}. 
The average integrated intensities in each subregion 
of key molecular species are summarized in Table \ref{region_integ}. 
We see the spectral lines of CCH, HCN, HCO$^+$, HNC, CS, SO, C$^{18}$O, 
and $^{13}$CO in the spectra of all the subregions, 
but with different relative intensities. 
In the subregion E, HNC, SO, and C$^{18}$O are faint, 
and the spectrum is rather similar to the spectrum averaged over the whole area. 
On the other hand, molecular species seen in the ``hot core'' spectrum, 
such as H$_2$CS, 
CH$_3$CCH, and OCS, are also detected in the spectrum of the subregion A. 
The lines of these species become fainter in the subregions B and C 
in this order, and are not detected in the subregion D. 

We closely compare the relative intensities of the lines 
of the key molecular species: CCH, HCN, HCO$^+$, HNC, C$^{34}$S, CH$_3$OH, 
CS, SO, HC$_3$N, C$^{18}$O, and $^{13}$CO, 
and investigated their variation among 5 subregions. 
For this comparison, we use the HCO$^+$ line, which is the second brightest 
after the $^{13}$CO line as a reference. 
It traces moderately denser area ($n_{H_2}\gtrsim10^4$ cm$^{-3}$) 
than the $^{13}$CO line, where the lines of the above molecules 
can be excited sub-thermally. 
As seen in Figure \ref{spectra_region}, 
the intensity ratios of almost all the species relative to HCO$^+$ 
become low from the subregion A to the subregion E in this order. 
The only exception is CCH. The CCH/HCO$^+$ ratio becomes higher 
in the subregion E. 
Note that the CS/HCO$^+$ and C$^{18}$O/HCO$^+$ ratios are 
highest at the subregion B. 
This may be caused by the line opacity 
and/or the chemical effect such as photodissociation near the star-forming core. 

The relatively high CCH/HCO$^+$ ratio in the outer subregions is obviously 
related to the widespread distribution of CCH. 
CCH is considered to be abundant in a photon dominated region (PDR). 
Since C$^+$ is rich in PDRs, CCH is efficiently 
produced from C$^+$ in the gas phase \citep{fuente2008chemistry}. 
PDRs are extended in cloud peripheries, and hence, 
CCH can be relatively abundant in the outer part of the cloud. 

In Figure \ref{spectra_region}, we find some differences 
in the intensities of HCN, HCO$^+$, HNC, and CS, among the 5 subregions, 
although they look similar in Figures \ref{map}, \ref{integint_correlation}, 
and \ref{integint_correlation_HCOp}. 
The intensity of these 4 species similarly behave in the subregions A--D. 
HCN, HNC, and CS lines become fainter relative to the HCO$^+$ line 
from the subregion A to the subregion D. 
In the subregion E, the intensity of the HNC line is fainter by a factor of 10 
than in the subregion D, while HCN, HCO$^+$, and CS lines are fainter 
by a factor of $5-7$ than in the subregion D. 
The excitation effect does not seem responsible 
for the faintness of HNC in the subregion E. 
The critical density ($n_{crit}$) of the HCN ($J=1-0$) line 
($1.0\times10^6$ cm$^{-3}$ at 10 K) is actually higher 
than that of the HNC ($J=1-0$) line ($2.8\times10^5$ cm$^{-3}$ at 10 K), 
as shown in Table \ref{critical}. 
Thus HNC is more deficient than HCN in the subregion E 
probably because of its formation and destruction processes. 
It is worth noting that HNC is less stable in energy than HCN 
by 0.64 eV \citep[KIDA database;][]{wakelam2012kida}, 
and hence, HNC is known to be isomerized to HCN 
under a high temperature condition \citep[$>24$ K;][]{hirota1998abundances}. 
Since the subregion E is expected to have higher temperature 
due to penetration of the interstellar UV radiation, 
the isomerization would occur efficiently, resulting in deficiency of HNC there. 

The detections of CCH, HCN, HCO$^+$, and CS in the subregion E 
indicate that the contribution of the extended gas 
may not be small for these molecular species in the averaged spectrum. 
They are also observed in diffuse clouds 
in absorption against extragalactic continuum sources 
\citep{lucas1996hcop, lucas2000cch, liszt2001hcn, lucas2002cs}. 
In the context of molecular-cloud-scale chemical compositions, 
\citet{pety2017anatomy} reported that 
the diffuse and translucent gas contributes to 
the 45\% or higher fluxes of CCH and HCO$^+$ 
in their 5.6 pc $\times$ 7.5 pc observation 
toward Orion B giant molecular cloud, while 
it contributes the $30-40$\% fluxes of HCN and CS. 
Although a precise comparison is difficult, 
our result seems qualitatively consistent with theirs. 
This result suggests that the contribution of the translucent gas 
should be evaluated statistically 
in various environments in order to extract information 
about the dense gas from molecular-cloud-scale observations. 
Some works in this direction have recently been reported 
\citep[e.g.,][]{shimajiri2017testing, kauffmann2017molecular, watanabe2017W51}. 

\begin{figure*}
\begin{center}
\includegraphics[width=\hsize, bb=0 0 556 552]{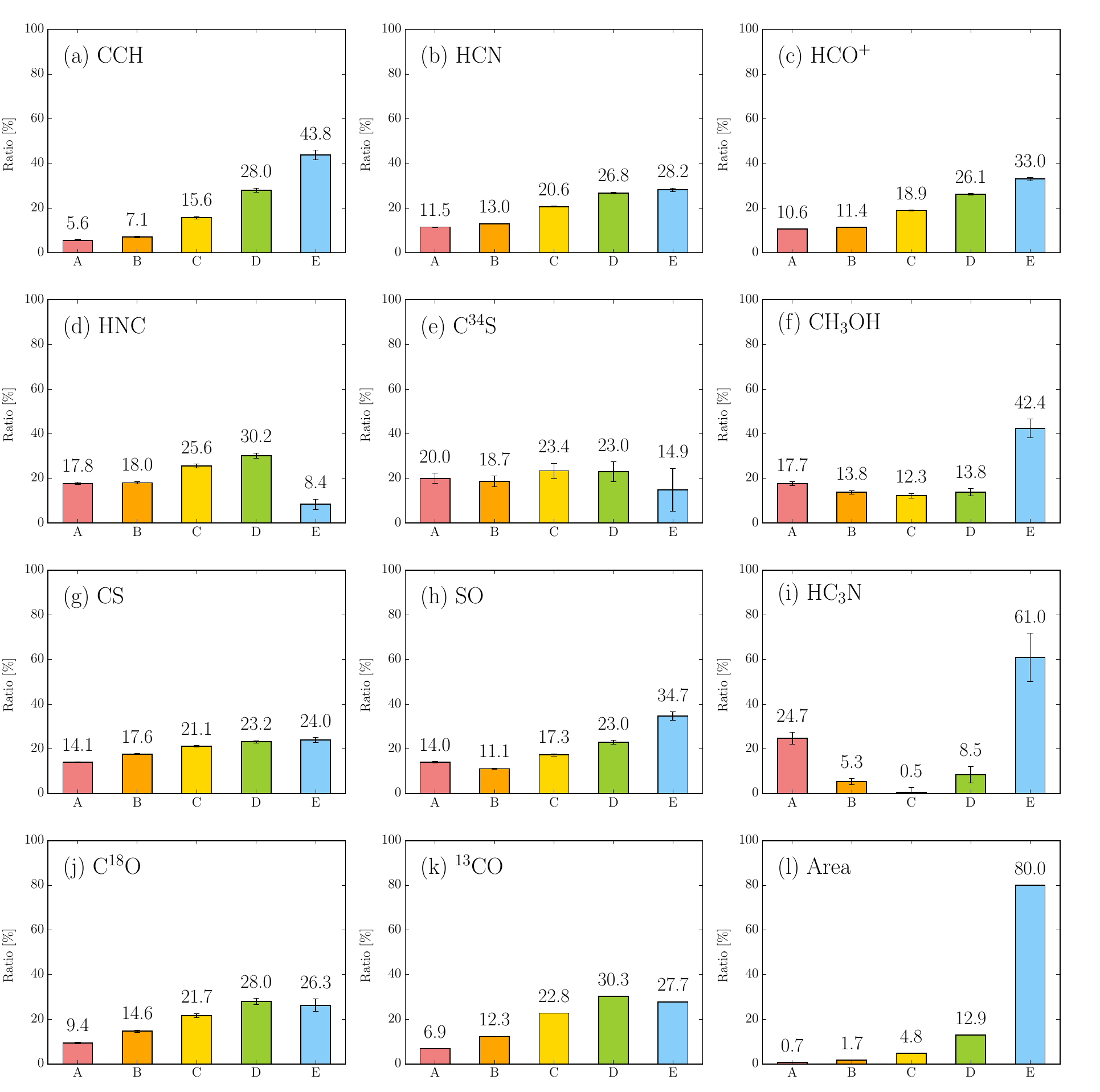}
\caption{(a)-(k) Fractional fluxes of each species 
and (l) fractional area for 5 subregions. 
Black lines indicate errors (1$\sigma$). 
A, B, C, D, and E are subregions classified 
by the $^{13}$CO integrated intensity. 
\label{pie}}
\end{center}
\end{figure*}

For a further analysis of the contribution of the extended gas, 
we prepare Figures \ref{pie} and \ref{fractionalflux}, 
which reveals a fraction of the flux of each subregion 
relative to the total flux of the spectrum averaged over the whole area. 
In Figure \ref{pie}, the fractional flux is defined as:  
$$
\textrm{Fractional Flux (\%)} =
\frac{\displaystyle\sum_{i\in \mathrm{R}}I_i(\mathrm{X})}
     {\displaystyle\sum_{i\in \mathrm{All}}I_i(\mathrm{X})}
\times 100\%,
$$
where $I_i(\mathrm{X})$ stands for the integrated intensity 
of a given molecule for the $i$-th $30''\times30''$ area and  
R represents the subregions A, B, C, D, and E. 
``All'' means the whole observed area. 
Likewise, the fractional area is defined as 
the number of the grids belonging to a given subregion 
divided by the total number of the grids. 
Figure \ref{fractionalflux} shows the spectral line profiles of 
individual molecular species, where the contribution from each subregion 
is indicated by colors. 
This figure is almost equivalent to Figure \ref{pie}, 
but better represents the contribution of each subregion 
including the uncertainties due to the signal-to-noise ratio of the spectra. 

In Figure \ref{pie}, we confirm that 
the contribution of the subregion A to the whole-area spectrum 
is small for all the molecular species. 
In particular, the flux of CCH mostly comes from the outer subregions D and E. 
It is surprising that more than 1/3 of CH$_3$OH and SO flux 
come from the subregion E, 
although the emissions of these species look concentrated around the hot core, 
as seen in Figure \ref{map}. 
Note that the large flux of HC$_3$N coming from subregion E 
(Figure \ref{pie}(i)) is likely affected by noise, 
as seen in Figure \ref{fractionalflux}(i). 
The low-level emissions of CH$_3$OH and SO extended 
in cloud peripheries are notable, 
because these species are often thought as good tracers of shocks 
associated with star formation. 
However, these molecules are also known to reside in cold dark clouds 
such as TMC-1 and L1544 
(\citealt{takakuwa1998h13cop} and 
\citealt{bizzocchi2014deuterated}, respectively), 
although their emissions are not as bright as those in star forming regions. 
Hence, accumulation of these weak emissions over a large area 
would make a significant contribution to 
the spectrum averaged over the whole area of W3(OH). 
This result gives an important implication for interpretation 
of the emissions of these molecules observed 
toward external galaxies in the 3 mm band. 

\begin{deluxetable}{cccc}
\tabletypesize{\scriptsize}
\tablecaption{Classification of subregions. \label{criteria}}
\tablewidth{0pt}
\tablehead{
\colhead{region} & \colhead{I($^{13}$CO)} 
& \colhead{number of data points} & \colhead{Area}\\
\colhead{} & \colhead{(K km s$^{-1}$)} & \colhead{} & \colhead{(arcmin$^{2}$)}}
\startdata
A & $>70$   &   7 &   1.75 \\
B & $50-70$ &  17 &   4.25 \\
C & $30-50$ &  49 &  12.25 \\
D & $10-30$ & 132 &  33.00 \\
E & $<10$   & 819 & 204.75 
\enddata
\end{deluxetable}

\begin{figure*}
\begin{center}
\includegraphics[width=0.6\hsize, bb=0 0 570 634]{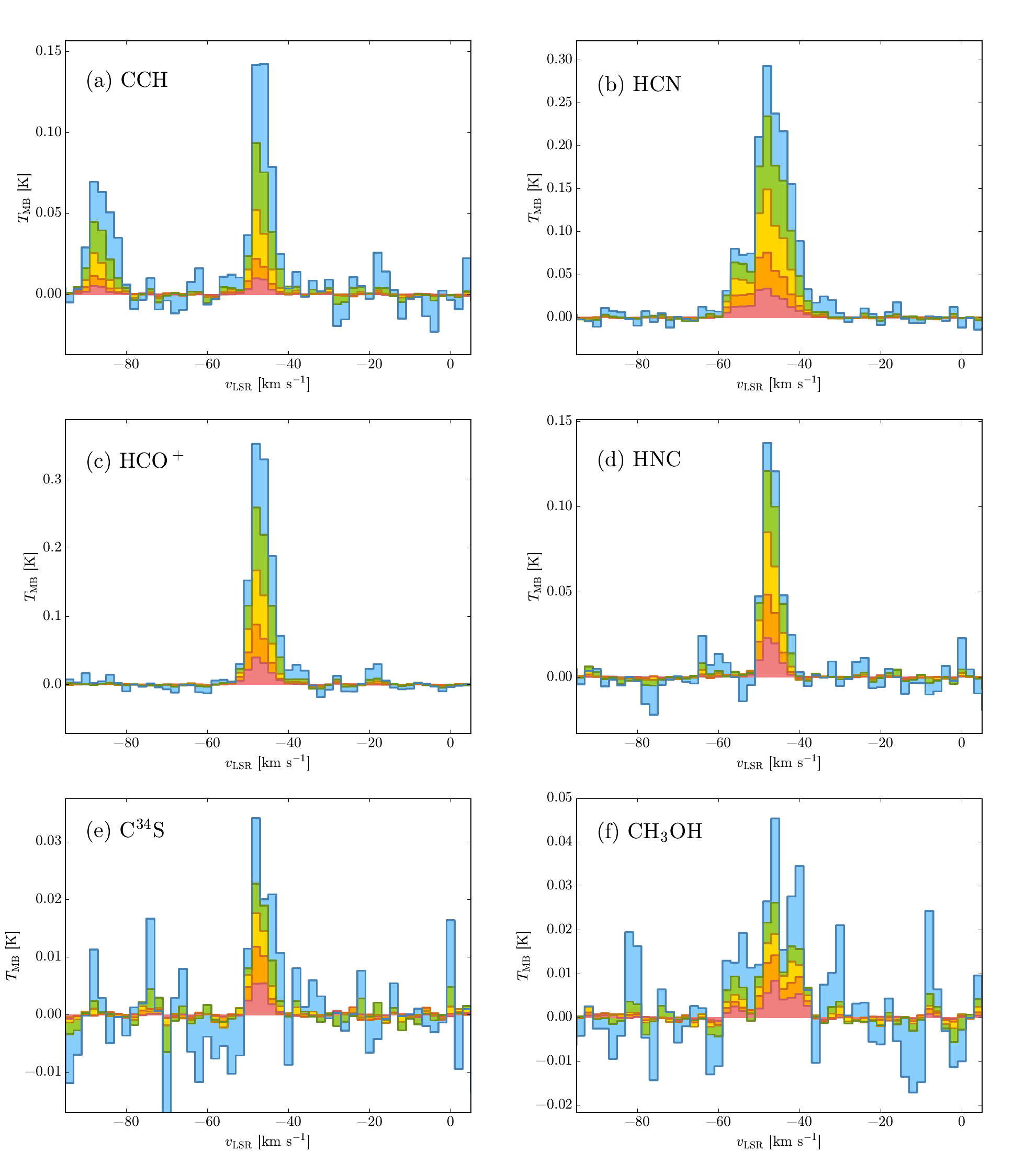}
\caption{Spectral line profiles of individual molecular transitions 
in the spectrum averaged over the whole area (9.0 pc $\times$ 9.0 pc). 
Contributions from each subregion are 
calculated by multiplication of ``flux from each subregion'' 
by ``fraction of the subregion in the whole area'' and are indicated by color. 
Note that the spectra are summed up from the inner subregion (A) 
to the outer subregion (E), namely, red is of subregion A, 
orange is of subregion A+B, yellow is of subregion A+B+C, and so forth. 
\label{fractionalflux}}
\end{center}
\end{figure*}

\addtocounter{figure}{-1}

\begin{figure*}
\begin{center}
\includegraphics[width=0.6\hsize, bb=0 0 570 634]{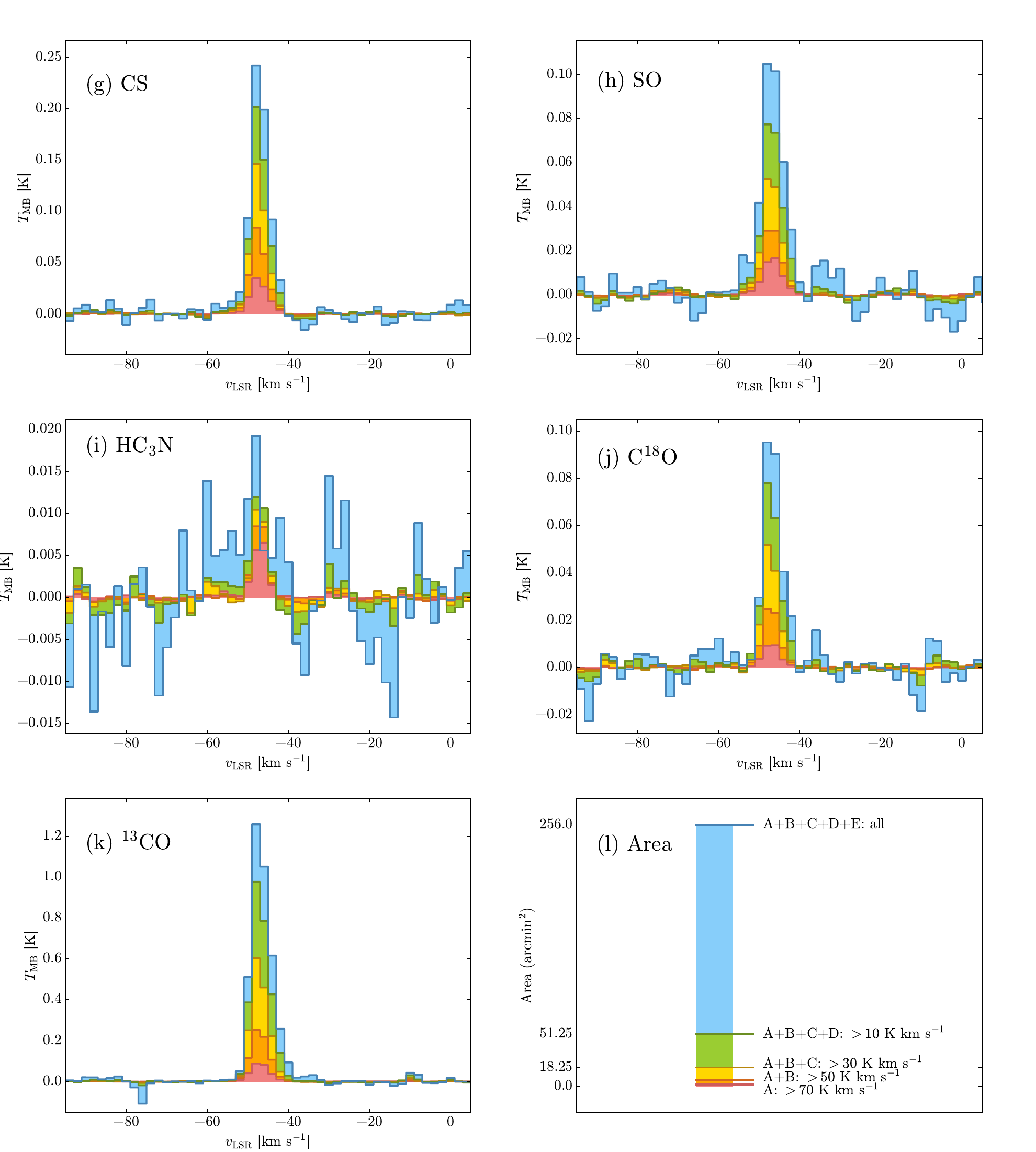}
\caption{\emph{(continued)}}
\end{center}
\end{figure*}

\subsection{Comparison with W51} \label{comparison}

In the first paper of this series \citep{watanabe2017W51}, 
we presented the mapping spectral line survey toward W51. 
W51 is a more active star-forming region than W3(OH). 
The cloud size of W51 is larger than that of W3(OH), 
and hence, the mapped area of W51, 39 pc $\times$ 47 pc, 
is larger than that for W3(OH) (9.0 pc $\times$ 9.0 pc). 
In spite of the large difference in the star-forming activity, the cloud size, 
and the location in the Milky Way, 
we find similarity of the spectra averaged over the observed area 
and the distributions of key molecular species between W3(OH) and W51. 

\begin{figure*}
\begin{center}
\includegraphics[width=\hsize, bb=0 0 721 432]{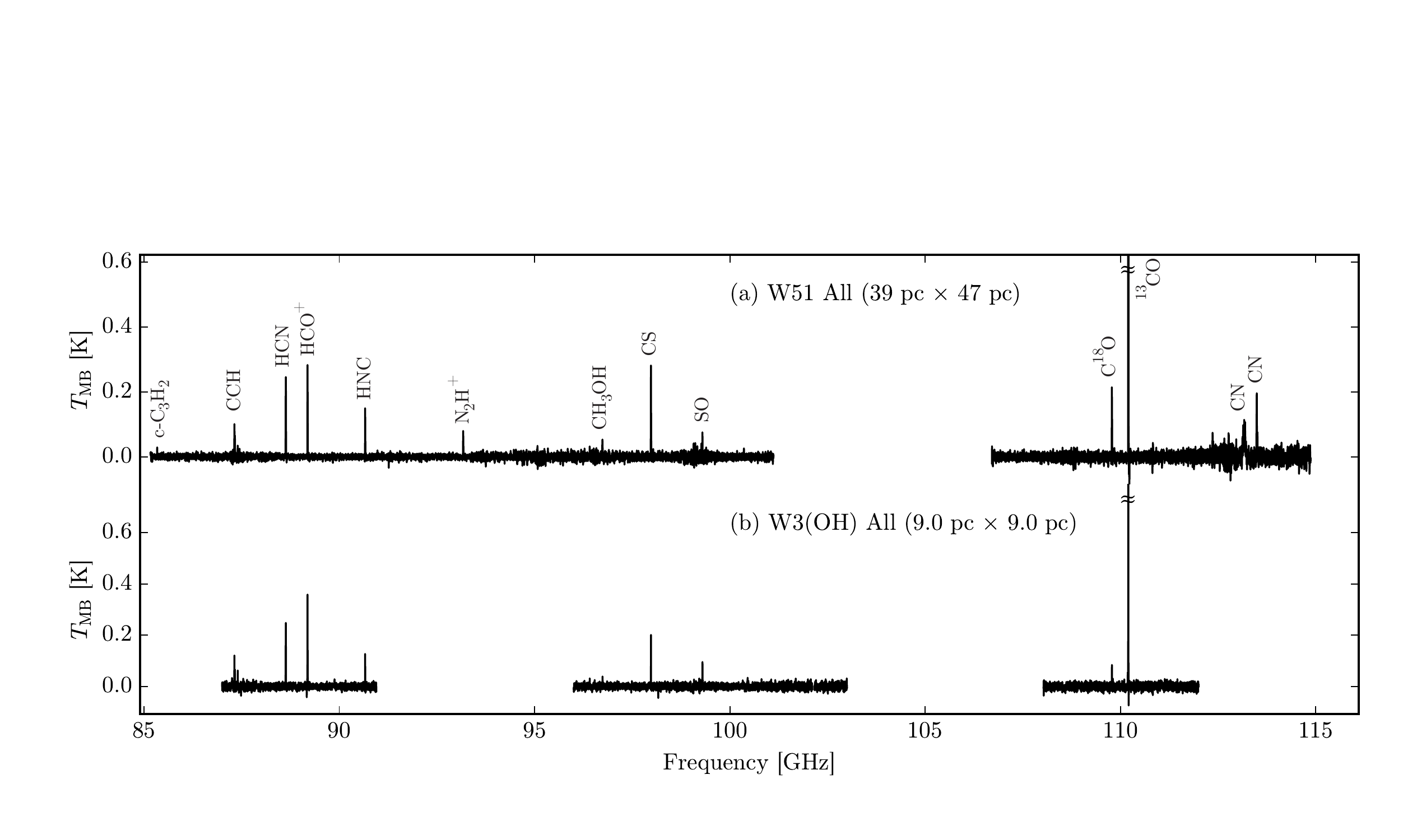}
\caption{Spectra of (a) a averaged over the observed area in W51 
presented by \citet{watanabe2017W51} 
and (b) spectrum averaged over the observed area in W3(OH) for comparison. 
\label{spectra_W51}}
\end{center}
\end{figure*}

\begin{figure}
\begin{center}
\includegraphics[width=\hsize, bb=0 0 432 432]{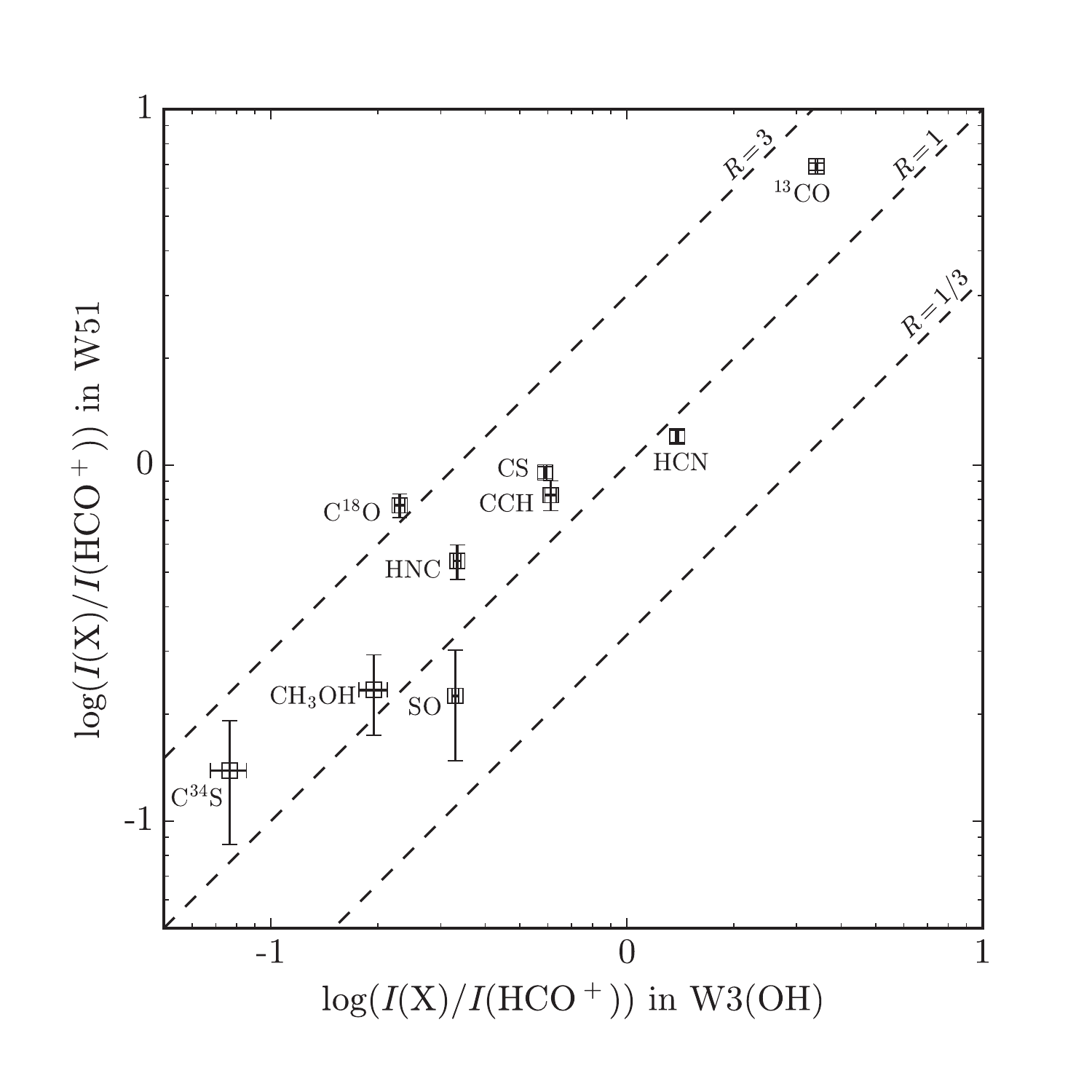}
\caption{Correlation diagram of the integrated intensity ratios 
relative to that of HCO$^+$ 
between W3(OH) (averaged over the 9.0 pc $\times$ 9.0 pc area) 
and W51 (averaged over the 39 pc $\times$ 47 pc area). 
The data for W51 is taken from \citet{watanabe2017W51}. 
Dashed lines indicate the ratios ($R$) of 3, 1, and 1/3. 
\label{W51}}
\end{center}
\end{figure}

Figure \ref{spectra_W51} shows 
the spectrum averaged over the whole area in W51. 
In the spatially averaged spectrum of W51, 
the lines of $c$-C$_3$H$_2$, CCH, HCN, HCO$^+$, HNC, N$_2$H$^+$, 
CH$_3$OH, CS, SO, C$^{18}$O, $^{13}$CO, and CN are detected. 
These molecular species are also identified 
in the spatially averaged spectrum of W3(OH) 
except for $c$-C$_3$H$_2$, N$_2$H$^+$ and CN, 
whose transitions are not covered in the present frequency settings. 
Furthermore, the spectral intensity pattern of W51 is 
similar to that of W3(OH) as a whole. 
Figure \ref{W51} shows a correlation diagram of integrated intensity ratios of 
CCH, HCN, HNC, C$^{34}$S, CH$_3$OH, CS, SO, C$^{18}$O, and $^{13}$CO 
relative to HCO$^+$ between W3(OH) and W51. 
Indeed, the correlation coefficient is 0.96 among 10 species: 
CCH, HCN, HCO$^+$, HNC, C$^{34}$S, CH$_3$OH, CS, SO, C$^{18}$O, and $^{13}$CO. 
Even if $^{13}$CO is excluded, it is still as high as 0.85. 

The similarity of the large-scale spectra of W51 and W3(OH) 
can further be confirmed, 
when we look at the spectra of the subregions, fractional fluxes, 
and correlation plots of the integrated intensities of molecules 
against the $^{13}$CO integrated intensity. 
The characteristics discussed in Section \ref{correlation} and \ref{subregion} 
are also the case in W51; 
CCH is relatively bright in the outer subregion of the cloud, 
while the other molecular species become faint relative to HCO$^+$ there. 

The minor difference is seen in the subregion spectra of HCN and C$^{18}$O. 
In the case of W51, their intensity ratios relative to HCO$^+$ are highest 
in the spectrum of the subregion C, 
while, in W3(OH), their intensity ratios decrease toward the outer subregions. 
The possible reason is the effect of optical thickness, 
the arbitrariness of the dividing criteria 
and/or photodissociation nearby the star-forming core. 
In spite of these small differences, we stress that the chemical composition 
characteristic to the local star-forming cores is mostly smeared out 
in the molecular-cloud-scale chemical composition observed in the 3 mm band. 
Therefore, molecular-cloud-scale chemical compositions for the both sources 
are essentially similar to each other. 
Moreover, this result means that no apparent effect of 
the different galactocentric distances for these two sources can be seen. 

\begin{deluxetable*}{lrlcccccc}
\tabletypesize{\scriptsize}
\tablecaption{Average integrated intensities of each subregion. 
\label{region_integ}}
\tablewidth{0pt}
\tablehead{
\colhead{Molecule} & \colhead{Frequency} & \colhead{Transition} & 
\colhead{A} & \colhead{B} & \colhead{C} & \colhead{D} & \colhead{E} & 
\colhead{9.0 pc $\times$ 9.0 pc averaged}\\
\colhead{} & \colhead{[GHz]} & \colhead{} & 
\colhead{(K km s$^{-1}$)} & \colhead{(K km s$^{-1}$)} & \colhead{(K km s$^{-1}$)} & 
\colhead{(K km s$^{-1}$)} & \colhead{(K km s$^{-1}$)} & \colhead{(K km s$^{-1}$)}}
\startdata
CCH & 87.316898 & $N=1-0$ & & & & & & \\[-1mm]
    &           & $J=5/2-3/2$ & & & & & & \\[-1mm]
    &           & $F=2-1$ &
$ 7.7 \pm 0.4 $ &
$ 4.0 \pm 0.2 $ &
$ 3.11 \pm 0.12 $ &
$ 2.06 \pm 0.07 $ &
$ 0.52 \pm 0.03 $ &
$ 0.95 \pm 0.03 $ \\
HCN & 88.631602 & $1-0$ &
$ 53.1 \pm 0.4 $ &
$ 24.7 \pm 0.3 $ &
$ 13.63 \pm 0.15 $ &
$ 6.57 \pm 0.09 $ &
$ 1.11 \pm 0.04 $ &
$ 3.16 \pm 0.03 $ \\
HCO$^+$ & 89.188525 & $1-0$ &
$ 36.9 \pm 0.3 $ &
$ 16.25 \pm 0.19 $ &
$ 9.37 \pm 0.11 $ &
$ 4.81 \pm 0.07 $ &
$ 0.98 \pm 0.03 $ &
$ 2.37 \pm 0.02 $ \\
HNC & 90.663568 & $1-0$ &
$ 18.5 \pm 0.3 $ &
$ 7.74 \pm 0.17 $ &
$ 3.81 \pm 0.10 $ &
$ 1.67 \pm 0.06 $ &
$ 0.08 \pm 0.02 $ &
$ 0.71 \pm 0.02 $ \\
C$^{34}$S & 96.412950 & $2_0-1_0$ &
$ 4.3 \pm 0.2 $ &
$ 1.66 \pm 0.15 $ &
$ 0.72 \pm 0.09 $ &
$ 0.26 \pm 0.05 $ &
$ 0.03 \pm 0.02 $ &
$ 0.148 \pm 0.019 $ \\
CH$_3$OH & 96.741375 & $2_0-1_0$, A$^+$ &
$ 10.9 \pm 0.3 $ &
$ 3.51 \pm 0.16 $ &
$ 1.08 \pm 0.10 $ &
$ 0.45 \pm 0.06 $ &
$ 0.22 \pm 0.02 $ &
$ 0.42 \pm 0.02 $ \\
CS & 97.980953 & $2-1$ &
$ 28.5 \pm 0.2 $ &
$ 14.69 \pm 0.15 $ &
$ 6.10 \pm 0.09 $ &
$ 2.49 \pm 0.06 $ &
$ 0.41 \pm 0.02 $ &
$ 1.38 \pm 0.02 $ \\
SO & 99.299870 & $N_J=2_3-1_2$ &
$ 15.4 \pm 0.2 $ &
$ 5.01 \pm 0.14 $ &
$ 2.71 \pm 0.09 $ &
$ 1.34 \pm 0.05 $ &
$ 0.33 \pm 0.02 $ &
$ 0.751 \pm 0.019 $ \\
HC$_3$N & 100.076392 & $11-10$ &
$ 4.9 \pm 0.2 $ &
$ 0.44 \pm 0.13 $ &
$ 0.01 \pm 0.08 $ &
$ 0.09 \pm 0.05 $ &
$ 0.104 \pm 0.019 $ &
$ 0.136 \pm 0.017 $ \\
C$^{18}$O & 109.782173 & $1-0$ &
$ 8.0 \pm 0.3 $ &
$ 5.09 \pm 0.16 $ &
$ 2.62 \pm 0.10 $ &
$ 1.26 \pm 0.06 $ &
$ 0.19 \pm 0.02 $ &
$ 0.58 \pm 0.02 $ \\
$^{13}$CO & 110.201354 & $1-0$ &
$ 78.4 \pm 0.3 $ &
$ 57.36 \pm 0.18 $ &
$ 36.80 \pm 0.10 $ &
$ 18.19 \pm 0.06 $ &
$ 2.68 \pm 0.03 $ &
$ 7.74 \pm 0.02 $ 
\enddata
\tablecomments{The velocity intervals of integration are 
from $-57$ to $-37$ km s$^{-1}$ except for HCN. \\
For HCN, the velocity range is from $-67$ to $-27$ km s$^{-1}$ 
to cover all the hyperfine components. }
\end{deluxetable*}

\subsection{Comparison with external galaxies} \label{comparison2}

Spectral intensity patterns of external galaxies 
observed by single-dish telescopes 
are also ones averaged over the molecular-cloud-scale. 
This is also true even for interferometer observations. 
The angular resolution readily achieved with the ALMA is about $0.2''$, 
which can be translated to the linear size of 10 pc 
for nearby galaxies at 10 Mpc. 
Hence, observed spectra toward external galaxies 
generally represent molecular-cloud-scale chemical composition. 

\begin{figure*}
\begin{center}
\includegraphics[width=\hsize, bb=0 0 864 332]{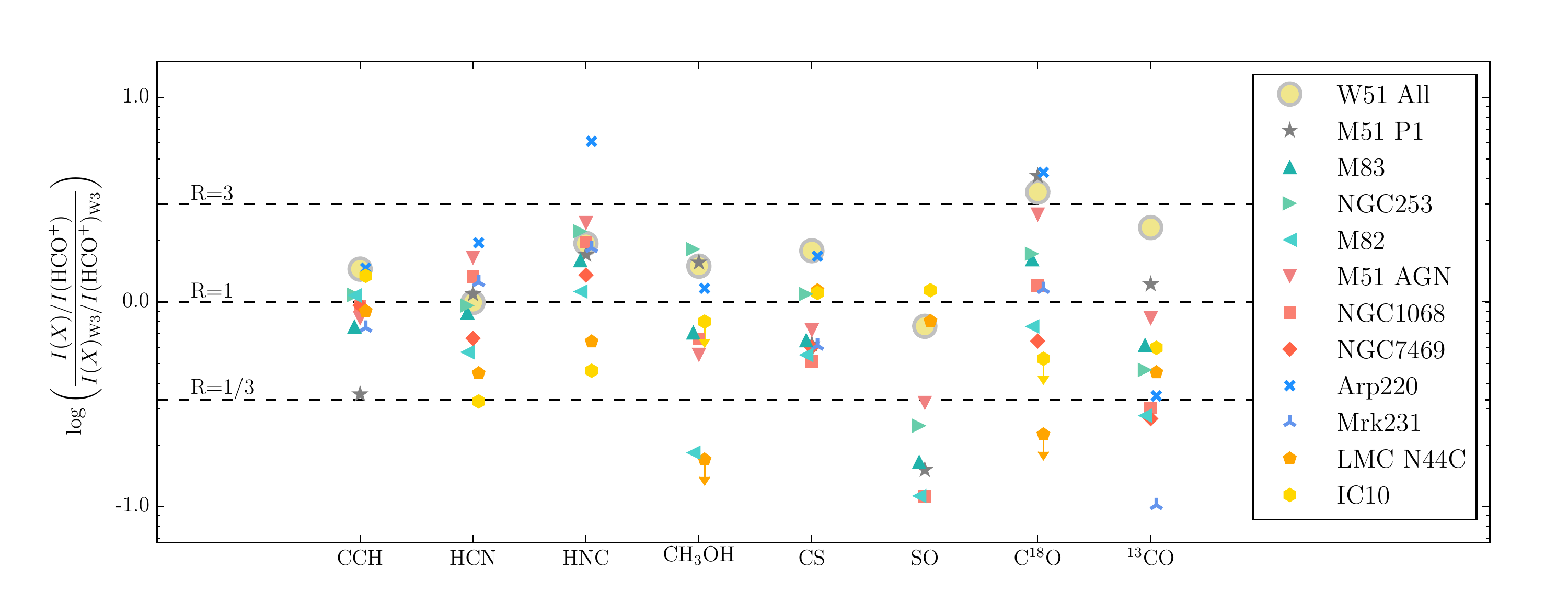}
\caption{Integrated intensity ratios of molecules 
relative to HCO$^+$ observed toward nearby galaxies. 
The values are normalized by that of W3(OH) 
(averaged over the 9.0 pc $\times$ 9.0 pc area). 
Dashed lines indicate the ratios ($R$) of 3, 1, and 1/3 relative to W3(OH). 
Integrated intensities are taken from the literature: 
W51 All (Galactic molecular cloud, 
averaged over the 39 pc $\times$ 47 pc area): \citet{watanabe2017W51};
M51 P1 (spiral arm): \citet{watanabe2014spectral}; 
LMC N44C: \citet{nishimura2016LMC}; 
IC10: \citet{nishimura2016IC10}; 
and others: \citet{aladro2015lambda}. 
\label{galaxies}}
\end{center}
\end{figure*}

We compare the integrated intensities of the bright molecular lines 
(CCH ($N=1-0$), HCN ($J=1-0$), HCO$^+$ ($J=1-0$), HNC ($J=1-0$), 
CH$_3$OH ($J_K=2_K-1_K$), CS ($J=2-1$), SO ($J_K=3_2-2_1$),  
C$^{18}$O ($J=1-0$), and $^{13}$CO ($J=1-0$)) observed in W3(OH) 
with those observed in nearby galaxies. 
Fortunately, a fairly complete dataset of these molecular lines 
observed toward various kinds of galaxies in the 3 mm band 
is available in the literature. 
Figure \ref{galaxies} shows the integrated intensity ratios 
of molecules relative to HCO$^+$ observed toward nearby galaxies. 
The values are normalized by that of the W3(OH) spectrum 
averaged over the 9.0 pc $\times$ 9.0 pc area. 
For this figure, we employ the data observed toward 
the spiral arm cloud (M51 P1), 
the central regions of the starburst galaxies (M83, NGC253, and M82), 
the active galactic nuclei (AGNs; M51 AGN, NGC1068, and NGC7469), 
the central regions of the ultra luminous infrared galaxies (ULIRGs; Arp220 and Mrk231) 
and the low-metallicity dwarf galaxies (the LMC N44C and IC10). 
The distances of these galaxies range from 50 kpc to 170 Mpc, 
and hence, the spatial resolutions range from 10 pc to a few kpc. 
For the most species, the intensity ratios agree with 
that of W3(OH) within a factor of 3, as seen in Figure \ref{galaxies}. 
This suggests that chemical compositions 
observed at a spatial resolution larger than 10 pc are generally similar, 
even if the spatial resolution is as large as a few kpc. 

In spite of the general resemblance, 
some molecular species in a specific galaxy 
seem to be influenced by its environment, as summarized below: 
\begin{itemize}
\item CCH: The CCH/HCO$^+$ ratio shows little variation among galaxies, 
and is close to the ratio for W3(OH). 
However, we note that the CCH/HCO$^+$ ratio seems to be enhanced 
in low-metallicity dwarf galaxies if the low elemental C/O ratio is considered 
\citep{nishimura2016IC10, nishimura2016LMC}. 
As mentioned before, CCH is efficiently produced in PDRs. 
Hence, galaxies hosting larger PDRs could be advantageously rich in CCH. 
However, no apparent trend suggesting this picture is seen between 
starburst galaxies, which should be dominated by PDRs, and AGNs. 
One possibility is that the spectra of AGNs may also be contaminated 
by starburst regions owing to the coarse observation beam. 
\item HCN and HNC: The HCN/HCO$^+$ and HNC/HCO$^+$ ratios show 
relatively small variations from those for W3(OH) 
except for the low-metallicity galaxies, where both HCN and HNC 
tend to be less abundant due to the lower elemental N abundance
\citep{nishimura2016IC10, nishimura2016LMC}. 
In Arp220, HNC is by far abundant. 
This is already discussed by \citet{aalto2007overluminous}, 
who suggest the intensity enhancement of HNC caused 
by pumping of the rotational levels of HNC through the mid-infrared continuum 
and/or effects of X-ray Dominated Regions (XDRs). 
\item CH$_3$OH: In the low-metallicity dwarf galaxies, 
CH$_3$OH is not detected (upper limits are shown in Figure \ref{galaxies}). 
The deficiency of CH$_3$OH in the dwarf galaxies is explained by 
the lower abundance of dust grains and the intense UV radiation due to it 
\citep{nishimura2016IC10, nishimura2016LMC}. 
In M82, CH$_3$OH is still detectable, but the intensity is weak. 
This is consistent with the fact that M82 is heavily influenced 
by the intense UV radiation \citep{fuente2008chemistry}. 
\item CS: The CS/HCO$^+$ ratio does not vary among galaxies, 
and is close to the ratio for W3(OH). 
Since the production and destruction processes of CS and HCO$^+$ 
are completely different, this trend may be fortuitous. 
However, this result might suggest the universal correlation 
between HCO$^+$ and CS in molecular-cloud-scale chemical composition. 
Further statistical studies are awaited. 
\item SO: The SO/HCO$^+$ ratios are lower 
except for the low-metallicity dwarf galaxies than in W3(OH). 
The spatial resolutions are 
10 pc and 80 pc for LMC N44C and IC10, respectively, 
while they are larger than several hundred pc for other galaxies. 
Hence, the SO/HCO$^+$ ratio may reflect 
the difference of a beam filling factor. 
It should be noted that SO is known to be enhanced by shocks 
due to outflows associated with protostars \citep[e.g.,][]{bachiller1997shock}. 
Such a contribution might affect molecular-cloud-scale spectra in the 3 mm band. 
Observations at a higher spatial resolution in the submillimeter-wave band 
will be useful to evaluate the contribution of such shocks. 
\item C$^{18}$O and $^{13}$CO: 
The ratio of C$^{18}$O/HCO$^+$ and $^{13}$CO/HCO$^+$ varies from galaxy to galaxy. 
In adittion, no apparent correlation between C$^{18}$O and $^{13}$CO can be seen. 
The effects of elemental isotope abundances, 
optical depth, and isotope selective photodissociation 
can be responsible for this feature. 
\end{itemize}

\section{Summary} \label{summary}
We have conducted a mapping spectral line survey for the 9.0 pc $\times$ 9.0 pc area of 
the Galactic molecular cloud W3(OH) with the NRO 45 m telescope.  
The observed frequency ranges are $87-91$, $96-103$, and $108-112$ GHz. 
The main results are summarized below. 

(1) Eight molecular species, CCH, HCN, HCO$^+$, HNC, CS, SO, C$^{18}$O, 
and $^{13}$CO, are identified in the spectrum 
averaged over all the observed area, 
while some molecular species such as OCS, H$_2$CS CH$_3$CCH, and CH$_3$CN 
detected in the hot core at a 0.17 pc resolution 
are missing in the averaged spectrum. 

(2) In the spatially averaged spectrum, 
emission of the species concentrated just around the star-forming core, 
such as CH$_3$OH is fainter than the hot core spectrum, 
whereas emission of the species widely extended over the cloud, 
such as CCH, is relatively stronger. 

(3) We have classified the observed area into the 5 subregions 
according to the integrated intensity of $^{13}$CO, 
and have evaluated the contribution of the molecular line flux of each subregion 
to the spectrum averaged over the whole observed area. 
It is suggested that the flux from the outer subregions 
(cloud peripheries) is not negligible 
even for the centrally concentrated species CH$_3$OH. 

(4) CCH, HCN, HCO$^+$, and CS are still strongly 
detected in the spectrum of the region 
with the $^{13}$CO integrated intensity lower than 10 K km s$^{-1}$. 
Hence, the contribution of the extended gas is indeed significant. 
The spectrum averaged over the molecular-cloud scale in the 3 mm band 
seems to represent the gas extended around the star-forming core, 
rather than star-forming regions. 

(5) The averaged spectrum is similar to the spectrum of 
the other Galactic molecular cloud W51 averaged 
over the 39 pc $\times$ 47 pc area, 
and also to the spectra observed in the external galaxies. 
The average spectrum can be regarded as the representative one, 
which is almost common in various molecular clouds. 

(6) Molecular line intensity ratios are compared with 
those for the Galactic molecular cloud W51 and the 10 external galaxies. 
Most intensity ratios agree with that of W3(OH) within a factor of 3. 
Especially, the CCH/HCO$^+$ and CS/HCO$^+$ ratios 
show little variation among galaxies. 
On the other hand, the CH$_3$OH/HCO$^+$ ratio 
tends to be low in dwarf galaxies and M82, 
probably due to the intense UV radiation. 
The C$^{18}$O/HCO$^+$ and $^{13}$CO/HCO$^+$ ratios 
seem to be affected by elemental isotope abundances, 
optical depth effects, and isotope selective photodissociation. 

\acknowledgments
We are grateful to the anonymous reviewer 
for valuable comments and suggestions. 
We thank the staff of the NRO 45 m telescope for excellent support.  
This study is partly supported from Grants-in-Aid of Education, Sports, 
Science, and Technologies of Japan (21224002, 25400223, and 25108005).  
Y.N. is supported by Grant-in-Aid for JSPS Fellows (268280).  

\software{NOSTAR, NEWSTAR}


\end{document}